%% file: main.tex
\documentclass[10pt,journal,compsoc]{IEEEtran}
\ifCLASSOPTIONcompsoc
  \usepackage[nocompress]{cite}
\else
  \usepackage{cite}
\fi
\ifCLASSINFOpdf
\else
\fi
\usepackage{float}

\usepackage{mathptmx} 

\newcommand{\ignore}[1]{}
\usepackage{fancyhdr}
\usepackage{wrapfig}

\usepackage[normalem]{ulem}
\usepackage[hyphens]{url}
\usepackage{hyperref}
 \usepackage{subfig}
\usepackage{url}
\usepackage{comment}
\usepackage{epsfig}
\usepackage[ruled,vlined,linesnumbered]{algorithm2e}
\usepackage{algorithmic}
\usepackage{booktabs, multirow} 
\usepackage{soul}
\usepackage[table]{xcolor} 
\usepackage{changepage,threeparttable} 

\usepackage{paralist}
\usepackage[colorinlistoftodos,prependcaption,textsize=small]{todonotes}
\presetkeys{todonotes}{disable,color=white}{}
\usepackage{booktabs}
\usepackage{tabularx}
\usepackage{amsmath}
\usepackage{amssymb}
\usepackage{setspace}
\usepackage{xspace}
\usepackage{rotating}
\usepackage{mathtools}
\usepackage{color}

\newcommand{\PopCorns}{Popcorns-Pro\xspace}
\begin{document}
%
\title{\Large\PopCorns: A Cooperative Network-Server Approach for Data Center Energy Optimization }
\author{\normalsize Sai Santosh Dayapule, Kathy Nguyen, Gregory Kahl,\\ 
	 Suresh Subramaniam, Guru Venkataramani\\
	Department of Electrical and Computer Engineering\\
	The George Washington University, Washington, DC, USA\\
	{\it \{saisantoshd, kathymn, gkahl, suresh, guruv\}@gwu.edu }
	}

\IEEEaftertitletext{\vspace{-2\baselineskip}}
\maketitle
\begin{abstract}
Data centers have become a popular computing platform for various applications,
and they account for nearly 2\% of total US energy consumption. Therefore, it
has become important to optimize data center power, and reduce their energy
footprint. With newer power-efficient designs in data center infrastructure and
cooling equipment,  active components such as servers and data center
networks consume a majority of power. Most
existing work optimize power in servers and networks independently, and do not
address them together in a holistic fashion that has the potential to achieve greater power
savings. In this article, we present \PopCorns, a cooperative server-network
framework for energy optimization. We present a comprehensive power model for heterogeneous data
center switches along with low power mode designs in combination with the
server power model. We design job scheduling algorithms that place tasks onto
servers in a power-aware manner, such that servers and network switches can take
effective advantage of low power states and available network link capacities.
Our experimental results show that we are able to achieve significantly higher
savings upto 80\% compared to the previously well-known server and network power
optimization policies.
\end{abstract}

%

\input{Introduction}
\input{systemdesign}

\input{problemstatement}
\input{algorithms}
\input{experiment}

\input{RelatedWork}

\input{Conclusion}
\input{acknowledgement}

\bibliographystyle{ieeetr}
{\bibliography{references}} 
%
%



\begin{wrapfigure}{l}{0.21\linewidth}
	\centering
		\includegraphics[width=\linewidth]{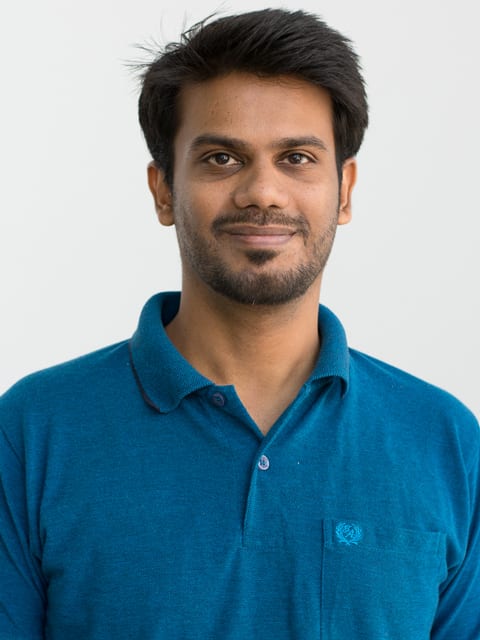}
\end{wrapfigure}
\textbf{Sai Santosh D} is currently pursuing the Ph.D. degree with the Department of Electrical and Computer Engineering, The George Washington University. His research 90oterests are in Datacenter energy Efficiency and computer architecture.

\bigbreak
\vspace{+10pt}

\textbf{Kathy Nguyen} is a graduate student with the Department of Electrical and Computer Engineering, The George Washington University. Her research interests are in Datacenter energy Efficiency and simulator development.

\bigbreak
\vspace{+10pt}

\textbf{Gregory Kahl} was an undergraduate student with the Department of Electrical and Computer Engineering, The George Washington University. His research interests are in Datacenter energy Efficiency and application development.

\bigbreak
\vspace{+10pt}

\begin{wrapfigure}{r}{0.21\linewidth}
	\centering
	\includegraphics[width=\linewidth]{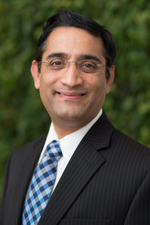}
\end{wrapfigure}
\textbf{Dr. Suresh Subramanian }is a Professor and a Chair of Electrical
and Computer Engineering with George Washington
University, Washington, D.C., where he directs the
Lab for Intelligent Networking and Computing. He
has published over 230 peer-reviewed papers in
these areas. His research interests are in the architectural, algorithmic, and performance aspects of
communication networks, with current emphasis on
optical networks, cloud computing, data center networks, and IoT.

\bigbreak
\vspace{+10pt}

\begin{wrapfigure}{l}{0.21\linewidth}
	\centering
	\includegraphics[width=\linewidth]{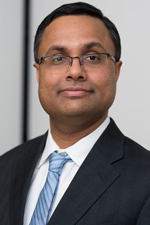}
\end{wrapfigure}

\textbf{Dr. Guru Venkataramani } is a Professor of Electrical
and Computer Engineering at George Washington
University, Washington, DC, USA. His research areas
are computer architecture, security, and energy
optimization. 
\end{document}

%% file: Introduction.tex
\section{Introduction}
\label{sec:Introduction}

Data centers have spurred rapid growth in computing, and an increasing number of user applications have continued to migrate toward cloud computing in the past few years. With this growing trend, data centers now account for about 2\% of US energy consumption \cite{barrosoCaseEnergyProportionalComputing2007}. Many public cloud computing environments have power consumption in the order of several Gigawatts. Therefore, energy is key challenges in data centers.

 Data center servers are typically provisioned for peak performance to always satisfy user demands. This, however, also translates to higher power consumption. Hardware investments have resulted in power saving mechanisms, such as DVFS and low-power or idle states~\cite{acpi}. 

We note that power reduction strategies in network switches and routers have been largely studied in large-scale network settings. Gupta et al.~\cite{guptaGreeningInternet2003} proposed a protocol-level support for coordinated entry into low-power states, where routers broadcast their sleep states for routing decisions to be changed accordingly. Adaptive Link Rate (ALR) for ethernet~\cite{gunaratneReducingEnergyConsumption2008} allows the network links to reduce their bandwidth adaptively for increased power efficiency. Such approaches may not be very effective in data center settings where application execution times have a higher dependence on network performance and Quality of Service (QoS) demands by the users. 

In this article, we propose PopCorns-Pro, a new framework to holistically optimize data center power through a cooperative network-server approach. We propose power models for data center servers and network switches (with support for low-power modes) based on power measurements in real system settings and memory power modeling tools from MICRON~\cite{lalSLCMemoryAccess2019}
and Cacti \cite{balasubramonianCACTINewTools2017}. We then study job placement algorithms that take communication patterns into account while optimizing the amount of sleep periods for both servers and line cards. Our experimental results show that we are able to achieve more than 20\% higher energy savings compared to a baseline strategy that relies on server load-balancing to optimize data center energy. 

We note that further power savings can be obtained at the application level through carefully tuning them for usage of processor resources~\cite{chenWattsinsideHardwaresoftwareCooperative2013} or through load-balancing tasks across cores in multicore processor settings to avoid keeping cores unnecessarily active. 
Such strategies can complement our proposed approach, and boost further power savings in data center settings.

 We extended upon our previous work~\cite{luPopCornsPowerOptimization2018} by proposing a multi-state power model for data center network switches and servers based on available power measurements in real system settings and memory power modeling tools. We formulate a power optimization problem that jointly considers both server
(with multiple cores) and network switches (with multiple line cards). We improve network traffic modeling accuracy by including link capacities. We also consider realistic heterogeneous switches in the Fat tree topology with different performance and power characteristics for switches at Core, Aggregate and Edge levels.
We compare our approach against a server load-balancing mechanism for task placement and a traditional Djikstra based network routing algorithm. We also consider a server energy optimization algorithm and a greedy bin packing network routing policy which consolidates traffic into fewer switches. These four policy combinations are compared with our Popcorns server algorithm to find that a combined server-network energy aware algorithm is more efficient than optimized versions of individual approaches previously considered as datacenter energy optimization. We consider real world job arrival traces, as well as with synthetic bursty arrivals at different job network demands to characterize the benefits of our combined server network selection policy in terms of energy consumption and job latencies. We found that our approach provides 25-80\% reduction in energy consumption compared to the case of optimizing server and network separately with conventional techniques.

 important goals and overall contributions of our work are:
\begin{itemize}
	\item We conceptualize the architectural sleep states for the switch, based on functional
	components and architectural design of real world data center switches. We
	motivate the benefit of such sleep states to the overall system power efficiency.
	\item We propose a new algorithm that considers servers and networks energy
	characteristics to coordinate server task placement while considering the power
	drawn by network components. Transition between power states in switches is
	controlled by buffer sizes and traffic patterns.
	\item We evaluate the impact of having the low-power states in the switch
	and our policy which optimizes for lower energy consumption in our data-center
	simulator. We consider a FatTree data center topology with various configuration parameters
	such as network traffic sizes, CPU traces, server and network performance
	models.
\end{itemize}

%% file: systemdesign.tex
\section{Power models}
\label{sec:systemdesign}
\label{sec:system-model}

\subsection{Conceptual Network Switch Low Power States}
\label{sec:switchpower}




\input{switchpower}

\subsection{Server Low-power States and Power Model}
\label{sec:serverpower}
With the focus on improving energy efficiency when servers are under-utilized, low-power states were created
with goal of reaching energy proportional computing. For standardization across computing platforms, Advanced Configuration
and Power Interface (ACPI)~\cite{acpi} specification gives the Operating system developers and hardware vendors common platform independent energy savings. 
ACPI has been historically well supported by various hardware vendors such as Intel and IBM~\cite{wareArchitectingPowerManagement2010}. ACPI uses global states, \emph{Gx}, to represent states of the
entire system that are visible to the user. We discuss the component-wise
breakdown of such system states as described in the table ~\ref{tab:serverpowermodel}.

Although processor sleep states can significantly reduce the power consumed by the processor, servers can still consume a considerable amount of power, as the platform may still remain active. As a result, in order to achieve further energy savings, \emph{system sleep states}, that also put platform components into low-power state, are considered for server farm power management. We consider additional full system sleep state. 


\begin{itemize}
	\item CPU(s) Power: Low-power states are now an important feature that are
	widely supported in today's processors. In a multicore processor, each core
	can have different architectural components or features disabled or turned off
	by shutting down the clock signal or turning off power and such states are
	called C-states. A higher level C state or S state typically indicates more
	aggressive energy savings but also corresponds to long wakeup latencies.
	For multi-core processor, low-power sleep states are supported at both core
	level and package level. When all cores become idle and reside in some
	\emph{Core C state}, the entire package would be resolved to a greater power saving state, denoted as \emph{package sleep state}, which further reduces power.

	\item{Chipset} The chipset and board consume energy used to support the main
	chipset, voltage regulators, bus control chips, interfaces for peripheral
	devices. According to Intel specifications~\cite{IntelX99Chipset} the chipset
	has TDP of 6.5W. The minimal power savings in the deeper sleep states of the
	system can be ignored.
	
	\item{DRAM}: The DRAM power consumption can be seperated into 3 parts ,
	read/write energy, Activate and pre-charge energy to open the specific row buffer to operate
	upon and refresh power to continuously perform the refresh cycle.
	Characterizing the DRAM power depends on several factors such as:
	
	1. DRAM technology DDR3, DDR4, LPDDR3L, etc with their respective power and
	performance characteristics.
	
	2. Workload : number of last level cache misses: An application whose working
	memory does not fit into the CPU cache can have a lot misses.
	
	3. Manufacturer optimizations: DRAM powerdown mode can be implemented
	differently by different manufacturers. The DRAM self-refresh cycle allows
	DRAM to be clock-gated from the external memory controller for self refreshing cycle.
	
	4. Memory clock speed. Higher clock improves performance buy consumes more memory.
	
	We used the micron power calculator to arrive the power consumption of the
	memory components as shown in the table ~\ref{tab:linecardpower}.
	
	\item{Power supply} We consider the AC-DC energy conversion loss which are
	typically 10\% of the power consumed by the system.
	
	\item{Fans}: The power consumption of a cooling fan is directly proportional to
	the cubic function of current fan speed utilization 
	
\end{itemize}

\begin{table*}
	\small	\centering
	\setlength{\extrarowheight}{0pt}
	\addtolength{\extrarowheight}{\aboverulesep}
	\addtolength{\extrarowheight}{\belowrulesep}
	\setlength{\aboverulesep}{0pt}
	\setlength{\belowrulesep}{0pt}
	\caption{A Representative Server Power model with Multi-socket Xeon processors and 128GB DDR3 }
	\label{tab:serverpowermodel}
	\resizebox{\linewidth}{!}{%
		\begin{tabular}{lcllll} 
			\toprule
			\textbf{Server Power Component } & \textbf{Number of units } & \textbf{Active } & \textbf{Idle (G0) } & \textbf{G1 } & \textbf{G2 } \\ 
			\hline
			\begin{tabular}[c]{@{}l@{}}\textbf{CPU}\\\textbf{(Intel Xeon E5}\\\textbf{V2 2690) }\end{tabular} & 2 & 135W & 108W & 22W & 15W \\
			\rowcolor[rgb]{0.851,0.851,0.851}  &  & \begin{tabular}[c]{@{}>{\cellcolor[rgb]{0.851,0.851,0.851}}l@{}}Average sustainable\\power consumption (TDP)\end{tabular} & \begin{tabular}[c]{@{}>{\cellcolor[rgb]{0.851,0.851,0.851}}l@{}}All CPUs/cores set\\to lowest DVFS \\frequency with\\20\% savings\end{tabular} & \begin{tabular}[c]{@{}>{\cellcolor[rgb]{0.851,0.851,0.851}}l@{}}All cores of all CPUs\\in C3 state- Clock\\Stopped and Cache\\flushed but other\\architectural state\\maintained\end{tabular} & \begin{tabular}[c]{@{}>{\cellcolor[rgb]{0.851,0.851,0.851}}l@{}}All cores of all CPUs\\in C6 state- Power\\gating the CPU, after\\saving the architectural\\state to the DRAM\end{tabular} \\
			\textbf{ Chipset } & 1 & 8W & 8W & 8W & 8W \\
			\rowcolor[rgb]{0.851,0.851,0.851}  &  & \begin{tabular}[c]{@{}>{\cellcolor[rgb]{0.851,0.851,0.851}}l@{}}Chipset powering \\the interfaces and\\peripherals\end{tabular} & \begin{tabular}[c]{@{}>{\cellcolor[rgb]{0.851,0.851,0.851}}l@{}}Chipset powering\\the interfaces and\\peripherals\end{tabular} & \begin{tabular}[c]{@{}>{\cellcolor[rgb]{0.851,0.851,0.851}}l@{}}Chipset powering the\\interfaces and\\peripherals\end{tabular} & \begin{tabular}[c]{@{}>{\cellcolor[rgb]{0.851,0.851,0.851}}l@{}}Chipset powering the\\interfaces and peripherals\end{tabular} \\
			\begin{tabular}[c]{@{}l@{}}\textbf{ Memory }\\\textbf{(8x16GB DDR3) }\end{tabular} & 8 & 1.45W & 0.29W & 0.29W & 0.29W \\
			\rowcolor[rgb]{0.851,0.851,0.851}  &  & \begin{tabular}[c]{@{}>{\cellcolor[rgb]{0.851,0.851,0.851}}l@{}}DRAM serving a normal\\utilization~of reads writes\\(25\% of all CPU~cycles\\each) + ACT~power (228mW) +\\I/0 power (540mw) + Background +\\Termination power(672mW)\end{tabular} & \begin{tabular}[c]{@{}>{\cellcolor[rgb]{0.851,0.851,0.851}}l@{}}Memory without new\\Reads or Writes, with ACT\\and background power\\consumption, and\\Background power 57mW.\end{tabular} & \begin{tabular}[c]{@{}>{\cellcolor[rgb]{0.851,0.851,0.851}}l@{}}Memory without new\\Reads or Writes, \\with ACT and\\background power consumption, and\\Background power 57mW.\end{tabular} & \begin{tabular}[c]{@{}>{\cellcolor[rgb]{0.851,0.851,0.851}}l@{}}Memory without new\\Reads or Writes, with\\ACT and background\\power consumption,\\and Background\\power 57mW.\end{tabular} \\
			\begin{tabular}[c]{@{}l@{}}\textbf{ Disks }\\\textbf{(8 TB SSDs}\\\textbf{in RAID) }\end{tabular} & 4 & 10W & 10W & 1W & 1W \\
			\rowcolor[rgb]{0.851,0.851,0.851}  &  & Active mode & Active mode & standby power & standby power \\
			\textbf{ Network Interface card } & 2 & 3W & 3W & 1W & 1W \\
			\rowcolor[rgb]{0.851,0.851,0.851}  &  & Active mode & Wake-on-LAN mode & Wake-on-LAN mode & Wake-on-LAN mode \\
			\textbf{ Power Supply losses } & 2 & 38 W & 31W & 7W & 5W \\
			\rowcolor[rgb]{0.851,0.851,0.851}  & \multicolumn{1}{l}{} & 10\% current system power consumption & 10\% current system power consumption & 10\% current system power consumption & 10\% current system power consumption \\
			\begin{tabular}[c]{@{}l@{}}\textbf{Cooling Fans}\\\textbf{15W fans}\end{tabular} & 2 & 10.5W & 5W & 5W & 0 W \\
			\rowcolor[rgb]{0.851,0.851,0.851}  &  & 70\% of Full Power & 30\% Power & 30\% Power & Off \\ 
			\hline
			\textbf{Total} & \multicolumn{1}{l}{} & \textbf{385W} & \textbf{308W} & \textbf{73W} & \textbf{~51W} \\ 
			\hline
			\textbf{ Transition Time - To sleep } &  & 0usecs & 10 usecs & 100 usecs & 500 msecs \\
			\rowcolor[rgb]{0.851,0.851,0.851}  &  &  & DVFS transition time & Transition to C3 state & \begin{tabular}[c]{@{}>{\cellcolor[rgb]{0.851,0.851,0.851}}l@{}}Saving Architectural state of CPU to ram\\and~power supply\\power up\end{tabular} \\ 
			\hline
			\textbf{ Wakeup latency } &  & 0 usecs & 100 usecs & 100 msecs & 1 second \\
			\rowcolor[rgb]{0.851,0.851,0.851}  &  &  & CPU frequency scale & Transiting all cores to C0 state & Transitioning CPU to Active state \\
			\bottomrule
		\end{tabular}
	}
\end{table*}

\subsection{Modeling Job}
More and more applications are being designed using a modular microservices-based paradigm, where inter-dependent tasks are hosted on different servers. The modularisation of job helps reduce complexities in software development, with independently update-able software programs in a scalable 'single-service instance per server' deployment pattern.  
We model the execution of jobs at the server side as consisting of multiple inter-dependent tasks that include both spatial and temporal inter-dependence. Application tasks are typically executed by specific server types. For example, a web service request will first be processed by an application or web server, and a search request is processed by a database server, and this kind of task relationship is called spatial inter-dependence. In terms of temporal inter-dependence, a task cannot start executing until all of its 'parent' tasks have finished their execution, and until after their results have been communicated to the server assigned to the task. A job is considered to have finished when all of its tasks finish execution. As for servers, there are multiple cores per server and one core can only process one task. We support asynchronous task execution by allowing the server running the parent task to  release the cpu, after all flows are completed to the child tasks' server even if it is waiting in the queue for execution.

\begin{figure}
	\captionsetup{font=small}
	\centering\includegraphics[width=0.4\linewidth]{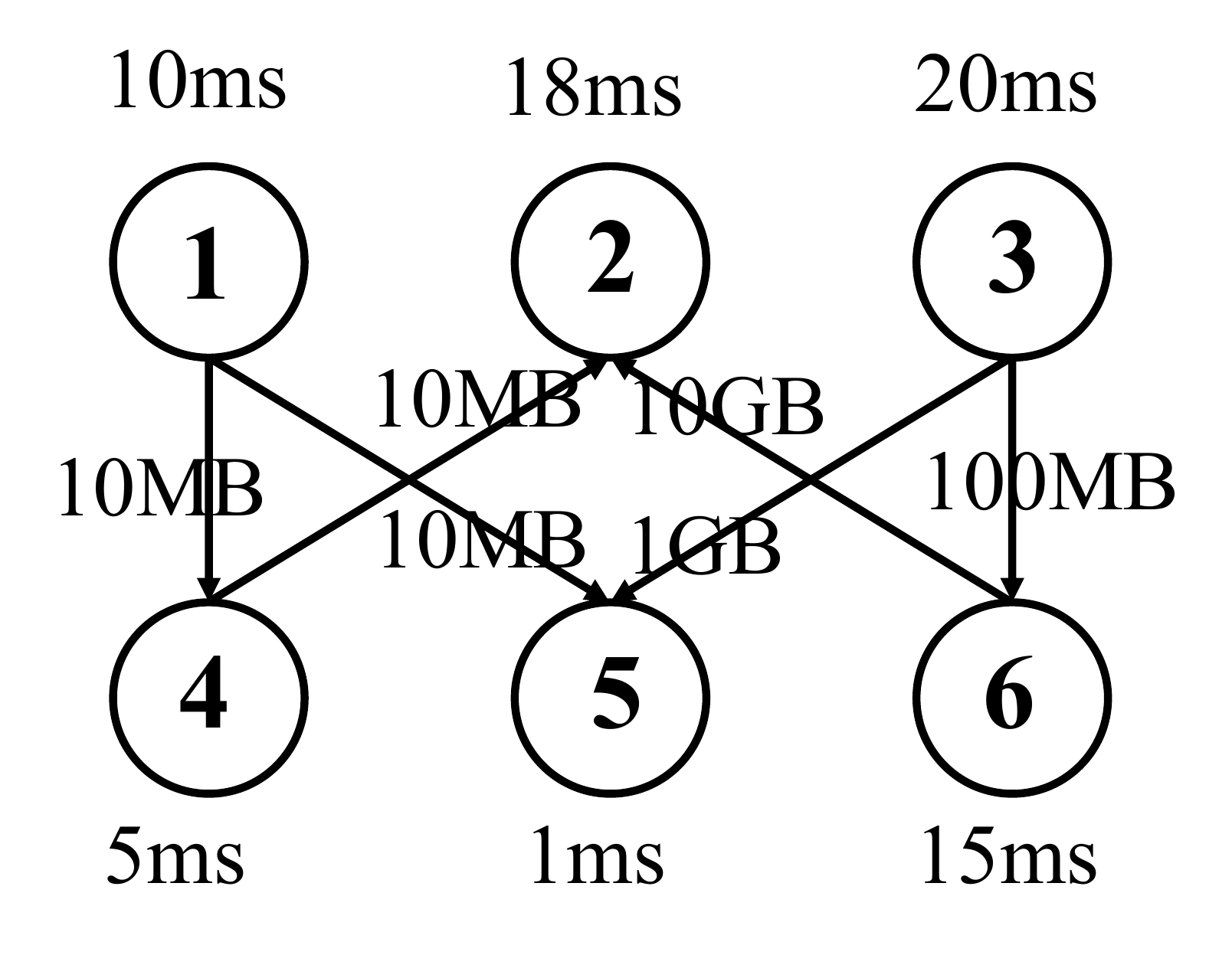}\caption{Example of a job DAG. Numbers 1-6 denote task 1-task 6 respectively. Numbers around the tasks represent task size, while numbers on the links represent flow size.}
	\label{figure1}
\end{figure}

Each job $j$ can be represented as a directed acyclic graph (DAG) $G^j(V^j, E^j)$, where $V^j$ is the set of tasks of job $j$. In DAG, if there is a link from task $i$ to task $r$, then task $i^j$ must finish and communicate its results to task $r^j$ before $r^j$ can start processing. Each task $v^j \in V^j$ has a workload requirement, namely task size or execution time requirement $w^j_v$ for the core. For each link in $E^j$, there is a data transfer size $D^j_l$ associated with it, which denotes the bandwidth requirement to transfer the result over link $l$ (from the task at the head of DAG link to the task at the tail) when assigned a network flow. Figure~\ref{figure1} shows an example of a job DAG.


\begin{figure}
	\captionsetup{font=small}
	\centering\includegraphics[width=0.3\textwidth]{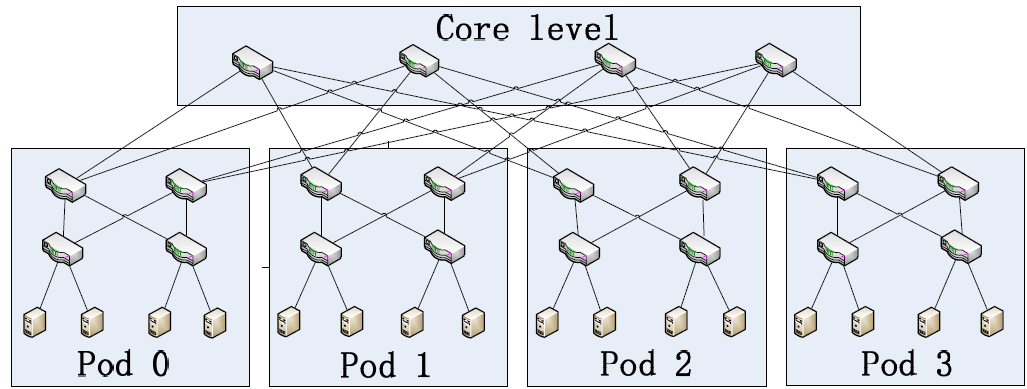}
	\caption{Fat tree topology with K=4 (16 servers).}
	\label{figure12}
\end{figure}

\subsection{Data Center System}
\label{sec:dc-sys}
Figure~\ref{figure12} shows the classic fat tree topology used in our network-server system, where each subset is called a 'pod'. In our system, each switch consists of a number of distributed cards plugged into the backplane, which provides the physical connectivity~\cite{panZerotimeWakeupLine2016}. Among these cards, there are multiple line cards for forwarding packets or flows, and can be in active, sleep, or off state. In turn, each line card contains several ports connecting to external links, which can also be in active, LPI, or off state. A typical schematic of switch, line card, and port is shown in Figure~\ref{figure13}.


%% file: switchpower.tex
Switch operates as OSI layer 2 network device, forwarding data frame from source
port to destination port and a router works at layer 3 for routing interdomain
network communication. A modular commercial switch such as the CISCO nexus line of switches consists of the following components:

\begin{itemize}
	\item \textit{Switch Chassis}: It houses the switch backplane, where all the
    modules are connected, power and cooling components. The chassis can also
    contain a variable number of configurable modules containing network
    interfaces and management modules.
	\item \textit{Modules}: There can two kinds: Supervisor and I/O modules. The supervisor modules contains the decision making components of the switch, and I/O modules such as linecard where network ports are connected.
\end{itemize}

Due to lack of detailed power models for commercial switches, we propose and
derive our switch power model based on available literature,
~\cite{panZerotimeWakeupLine2016} and corroborating with the cisco power
calculator~\cite{CiscoPowerCalculator} for Nexus line of switches, and tools
such as Cacti ~\cite{balasubramonianCACTINewTools2017}  Micron power modeling tools~\cite{MicronMemoryPower}.

\subsubsection{Supervisor modules}
\begin{figure}
	\captionsetup{font=small}
	\centering\includegraphics[width=1.2\linewidth]{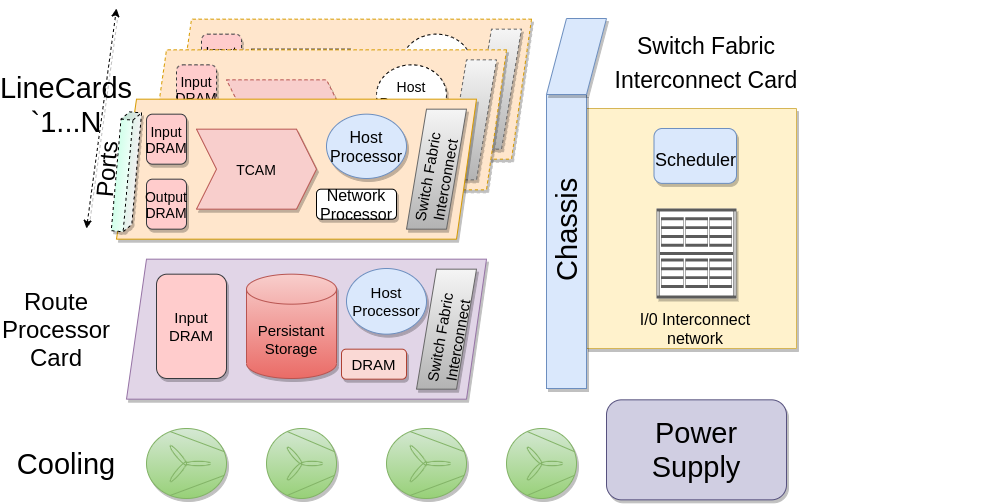}\caption{Illustration of Switch architecture.}	\label{figure13}
\end{figure}

As shown in figure ~\ref{figure13}, the control pane where all
forwarding decisions is performed by the SUP(Supervisor) module which consists
of the Route Processor (OSI layer 3 function) and Switch Processing card(OSI
layer 2 functions).
The CISCO SUP2E contains DRAM upto 32GB(Core switches).
According to Cisco power calculator~\cite{CiscoPowerCalculator}, a routing
processor card consumes upto 69W. We breakdown the components of the route
processing card in table ~\ref{tab:switchchassis}.

\subsubsection{Line Card power Model}
A linecard consists of several components such as ASICs, TCAMs, DRAM memory, and ports. Our power model for each component is explained below:

Due to lack of detailed power models for commercial switches, we propose and
derive our switch power model based on available literature and memory power
modeling tools from Micron~\cite{MicronMemoryPower}. A switch used in production uses
the several components such as ASICs, TCAMs, DRAM memory, and ports. Our power model for each component is explained below:

\label{sec:switcharch}

\begin{table*}
	\centering
	\small
	\setlength{\extrarowheight}{0pt}
	\addtolength{\extrarowheight}{\aboverulesep}
	\addtolength{\extrarowheight}{\belowrulesep}
	\setlength{\aboverulesep}{0pt}
	\setlength{\belowrulesep}{0pt}
	\caption{Single Linecard Power model for an Edge level Switch with 1 Gbps maximum bandwidth}
	\label{tab:linecardpower}
	\resizebox{\linewidth}{!}{%
		\begin{tabular}{llllll} \toprule
			& \multicolumn{1}{c}{\begin{tabular}[c]{@{}c@{}}\textbf{Active }\\\textbf{(1 Gbps~Bandwidth )}\end{tabular}} & \textbf{Low power state 1} & \textbf{Low power state 2} & \textbf{Low Power state 3} & \begin{tabular}[c]{@{}l@{}}\textbf{Line Card~ -}\\\textbf{Deepest Sleep State~}\end{tabular} \\ \hline
			\begin{tabular}[c]{@{}l@{}}Packet Forwarding\\~(Forwarding and \\replication Engines)\end{tabular} & 165W & 132W & 66W & 33W & 0W \\
			\rowcolor[rgb]{0.882,0.882,0.882}  & \begin{tabular}[c]{@{}>{\cellcolor[rgb]{0.882,0.882,0.882}}l@{}}Active-Max Clock\\frequency\end{tabular} & \begin{tabular}[c]{@{}>{\cellcolor[rgb]{0.882,0.882,0.882}}l@{}}DVFS savings for\\the network processor.\end{tabular} & \begin{tabular}[c]{@{}>{\cellcolor[rgb]{0.882,0.882,0.882}}l@{}}Clock gating the\\lookup~caches in\\the forwarding\\engines, Replication\\engines clock gated.\end{tabular} & \begin{tabular}[c]{@{}>{\cellcolor[rgb]{0.882,0.882,0.882}}l@{}}Forwarding engine's \\Core clock and bus \\stopped.\end{tabular} & \begin{tabular}[c]{@{}>{\cellcolor[rgb]{0.882,0.882,0.882}}l@{}}Architectural state\\flushed to DRAM\\and Forwarding\\engine and replication\\engines are power gated.\end{tabular} \\
			\begin{tabular}[c]{@{}l@{}}VoQ (M)\\(INPUT + OUTPUT)(max)\end{tabular} & \begin{tabular}[c]{@{}l@{}}15.5W +\\2 x 6 mW per MB of flow\end{tabular} & \begin{tabular}[c]{@{}l@{}}15.5W+\\1 x 6mW per MB flow\end{tabular} & 4W & 0 W & 0 W \\
			\rowcolor[rgb]{0.882,0.882,0.882}  & \begin{tabular}[c]{@{}>{\cellcolor[rgb]{0.882,0.882,0.882}}l@{}}Both Input and\\Output buffers active\end{tabular} & \begin{tabular}[c]{@{}>{\cellcolor[rgb]{0.882,0.882,0.882}}l@{}}Output VoQ buffer\\turned off\end{tabular} & \begin{tabular}[c]{@{}>{\cellcolor[rgb]{0.882,0.882,0.882}}l@{}}DRAM clock gated\\and contents lost.\end{tabular} & DRAM Power gated & DRAM power gated \\
			TCAM(max) & \begin{tabular}[c]{@{}l@{}}12W+ \\2.6 mW per MB of flow\end{tabular} & \begin{tabular}[c]{@{}l@{}}12W+2.6 mW per MB\\of flow\end{tabular} & \begin{tabular}[c]{@{}l@{}}12W+2.6 mW per MB\\of flow\end{tabular} & 12W & 0W \\
			\rowcolor[rgb]{0.882,0.882,0.882}  & \begin{tabular}[c]{@{}>{\cellcolor[rgb]{0.882,0.882,0.882}}l@{}}TCAM active and\\synchronized\end{tabular} & \begin{tabular}[c]{@{}>{\cellcolor[rgb]{0.882,0.882,0.882}}l@{}}TCAM Active and\\synchronized\end{tabular} & \begin{tabular}[c]{@{}>{\cellcolor[rgb]{0.882,0.882,0.882}}l@{}}TCAM active and\\synchronized\end{tabular} & \begin{tabular}[c]{@{}>{\cellcolor[rgb]{0.882,0.882,0.882}}l@{}}TCAM stopped,\\synchronized,\\and clock gated.\end{tabular} & \begin{tabular}[c]{@{}>{\cellcolor[rgb]{0.882,0.882,0.882}}l@{}}TCAM-routing\\tables flushed and\\power gated\end{tabular} \\
			Interconnect Fabric interface & 23W & 23W & 23W & 0 W & 0 W \\
			\rowcolor[rgb]{0.882,0.882,0.882}  & \begin{tabular}[c]{@{}>{\cellcolor[rgb]{0.882,0.882,0.882}}l@{}}Active for Control\\plane operations\end{tabular} & \begin{tabular}[c]{@{}>{\cellcolor[rgb]{0.882,0.882,0.882}}l@{}}Active for Control\\plane operations\end{tabular} & \begin{tabular}[c]{@{}>{\cellcolor[rgb]{0.882,0.882,0.882}}l@{}}Active for Control\\plane operations.\end{tabular} & \begin{tabular}[c]{@{}>{\cellcolor[rgb]{0.882,0.882,0.882}}l@{}}Power gated,\\Control plane\\operation stopped\end{tabular} & \begin{tabular}[c]{@{}>{\cellcolor[rgb]{0.882,0.882,0.882}}l@{}}Power gated,\\Control plane\\operation stopped\end{tabular} \\
			Host Processor & 24 W & 22 W & 9 W & 9 W & 3 W \\
			\rowcolor[rgb]{0.882,0.882,0.882}  & \begin{tabular}[c]{@{}>{\cellcolor[rgb]{0.882,0.882,0.882}}l@{}}Active and running\\Linecard OS\end{tabular} & \begin{tabular}[c]{@{}>{\cellcolor[rgb]{0.882,0.882,0.882}}l@{}}Active and running\\Linecard OS, with\\DVFS\end{tabular} & C3 state. Halt mode. & C3 state. Halt mode & \begin{tabular}[c]{@{}>{\cellcolor[rgb]{0.882,0.882,0.882}}l@{}}C5 state: deeper\\sleep state\end{tabular} \\
			Ports & 29 W & 15W & 4 W & 4 W & 4 W \\
			\rowcolor[rgb]{0.882,0.882,0.882}  & All 24 ports active & \begin{tabular}[c]{@{}>{\cellcolor[rgb]{0.882,0.882,0.882}}l@{}}Output ports in Low\\Power Idle- Wake on Arrival\end{tabular} & \begin{tabular}[c]{@{}>{\cellcolor[rgb]{0.882,0.882,0.882}}l@{}}All ports in Low\\Power Idle-Wake\\on Arrival\end{tabular} & \begin{tabular}[c]{@{}>{\cellcolor[rgb]{0.882,0.882,0.882}}l@{}}All ports in Low\\Power Idle- Wake on Arrival\end{tabular} & \begin{tabular}[c]{@{}>{\cellcolor[rgb]{0.882,0.882,0.882}}l@{}}All ports in Low\\Power Idle- Wake on Arrival\end{tabular} \\ \hline
			\begin{tabular}[c]{@{}l@{}}\textbf{Max Power Consumption~~}\\\textbf{in each state (}\\\textbf{1 Gbps when active)}\end{tabular} & \textbf{310 W} & \textbf{250 W} & \textbf{151 W} & \textbf{45 W} & \textbf{7 W} \\ \bottomrule
		\end{tabular}
	}
\end{table*}

\begin{table}
	\centering
	\scriptsize
	\caption{Switch Chassis power model.}
	\label{tab:switchchassis}
	\begin{tabular}{lll} 
		\toprule
		\multicolumn{1}{c}{\begin{tabular}[c]{@{}c@{}}\textbf{Switch Chassis }\\\textbf{components}\end{tabular}} & \multicolumn{1}{c}{\begin{tabular}[c]{@{}c@{}}\textbf{At-least}\\\textbf{One}\\\textbf{Line card~}\\\textbf{Active}\end{tabular}} & \multicolumn{1}{c}{\begin{tabular}[c]{@{}c@{}}\textbf{All Line }\\\textbf{Cards}\\\textbf{in OFF}\\\textbf{state}\end{tabular}}  \\ 
		\hline
		\multicolumn{3}{c}{\textbf{Route processing card}}                                                                                                                                                                                                                                                                                                                               \\ 
		\hline
		Hostprocessor                                                                                             & 24W                                                                                                                               & 0                                                                                                                                \\
		DRAM-3GB                                                                                                  & 32.5W                                                                                                                             & 0                                                                                                                                \\
		Persistent Storage                                                                                        & 10W                                                                                                                               & 0                                                                                                                                \\ 
		\hline
		\multicolumn{3}{c}{\textbf{\textbf{Switch Interconnect card}}}                                                                                                                                                                                                                                                                                                                   \\ 
		\hline
		Switch Fabric and~~I/0 (40\%load)                                                                         & 85W                                                                                                                               & 0                                                                                                                                \\
		Scheduler ASIC                                                                                            & 24W                                                                                                                               & 0                                                                                                                                \\ 
		\hline
		\multicolumn{3}{c}{\textbf{Chassis Controller}}                                                                                                                                                                                                                                                                                                                                  \\ 
		\hline
		Host-Processor                                                                                            & 24W                                                                                                                               & 24W                                                                                                                              \\
		DRAM-3GB                                                                                                  & 12W                                                                                                                               & 12W                                                                                                                              \\ 
		\hline
		\multicolumn{3}{c}{\textbf{Power Proportional Chassis Components}}                                                                                                                                                                                                                                                                                                               \\ 
		\hline
		Power supply losses~per Active card (25\%)                                                                & 93W                                                                                                                               & 5W                                                                                                                               \\
		Cooling per Active Line~\textasciitilde{}Card (20\%)                                                      & 77W                                                                                                                               & 4W                                                                                                                               \\
		\bottomrule
	\end{tabular}
\end{table}

\textbf{1. ASICs/ Network Processors:} The data plane operations of the switch such as
parsing the packet contents to read the header, looking up routing tables, and
forwarding data to the corresponding destination port are performed by the
networking processor. The processing is divided by two functional components:
Forwarding Engine and Replication Engine. The Replication engine separates the
header and data parts of a network packet and passes the header to the forwarding
engine. The replication engine also reassembles the packet after its processed
by the forwarding engine and in multicast communication duplicate packets to
different destination ports. The forwarding engine is the decision making
component in the line cards, and it uses the locally synchronized routing
tables, QoS and ACL lookups. According to Wobker's report \cite{wobkerPowerConsumptionHighend2012}, this
consumes 52\% of the total power in enterprise Cisco line cards.  Accordingly,
the ASIC/network processor's power consumption is computed to be 165 W. Based on
studies done by Iqbal et al. \cite{iqbalEfficienttrafficaware2012} and Luo et
al.

To construct the energy saving sleep states, we draw parallels to the design
of the sleep states in general purpose processors. In the first idle state, the
clock can be set to the lowest frequency therby resulting in 20\% savings. In
the first sleep state, the lookup caches in the forwarding engines can be flushed
and clock gated along with the replication engines. In the next sleep state, the core processing clock and processor bus is clockgated.
In the lowest sleep state state, the entire architectural state of the processor including runtime data from Fowarding information base rules, Qos Classification counters is written to the Linecard DRAM and power gated.

\textbf{2. VoQ (DRAM memory and SRAM buffer):} In Cisco line cards, the Virtual Output
Queuing memory is used to provide buffering and queing function to manage the
QoS, packet replication, congestion control functions. The ingress DRAM is used to buffer incoming
packets and the egress DRAM buffer is used store data payload while the header
is being processed by the forwarding engine.  The active power consumption of
DRAM depends on the frequency of accesses, and leakage/static power depends on
the transistor technology. Micron Power calculators \cite{MicronMemoryPower} for
RLDDR3 (Reduced latency DRAM) show a power consumption of 1571 mW per 1.125 GB
of memory when active and 314 mW when in sleep.
There is also a high speed SRAM buffer which acts as a write-back cache for the
DRAM memory.

\textbf{3. TCAM (Ternary Content addressable memory):} The TCAM structre is used by the
L3 routing function and to store routing table entries synchronized locally from
the route processing card. A typical 4.5 Mb TCAM structure which is used to offload high-speed packet lookup, consumes 15 W of power \cite{guoResistiveTCAMAccelerator2011}. We model the static leakage power for a 4.5 Mb CAM structure using Cacti \cite{balasubramonianCACTINewTools2017}, which estimates the power consumed during the idle sleep period when memory is not accessed.

\textbf{4. Line card Interconnect fabric:} The line card communicates with the chassis
bus using the interconnect interface. The line card fabric interconnect consumes 23W during active power state \cite{panZerotimeWakeupLine2016}. 

\textbf{5. Host processor and local DRAM:} Each line card includes a host processor which is used in line card boot and initialization process for copying routing table information from the switch fabric card. The processor is kept running in sleep mode to keep the routing tables synchronized and to wake up the line card on packet arrival. We assume a 30\% power reduction due to dynamic frequency scaling during line card sleep \cite{liuSleepScaleRuntimeJoint2014}.

\textbf{6. Ports:} To tackle the energy consumption in network equipment, the IEEE 802.3az standard introduces the Low Power Idle (LPI) mode of Ethernet ports, which is used when there is no data to transmit, rather than keeping the port in active state all the time~\cite{gunaratneReducingEnergyConsumption2008}. The idea behind LPI is to refresh and wake up the port when there is data to be transmitted; the wakeup duration is usually small. 

Apart from the line cards, we model a constant baseline power of 120 W for the rest of the switch in ON state, which includes switch supervisor module, the backplane, cooling systems, switch fabric card based on Pan et al. \cite{panZerotimeWakeupLine2016}.  The wakeup latency for each of the states is derived from sleep and wake up state conceptualized in section~\ref{sec:transition}

\subsection{Modeling switch sleep state transition and wake up latency}

\label{sec:transition}
\begin{figure}
\captionsetup{font=small}
\centering\includegraphics[width=.5\textwidth]{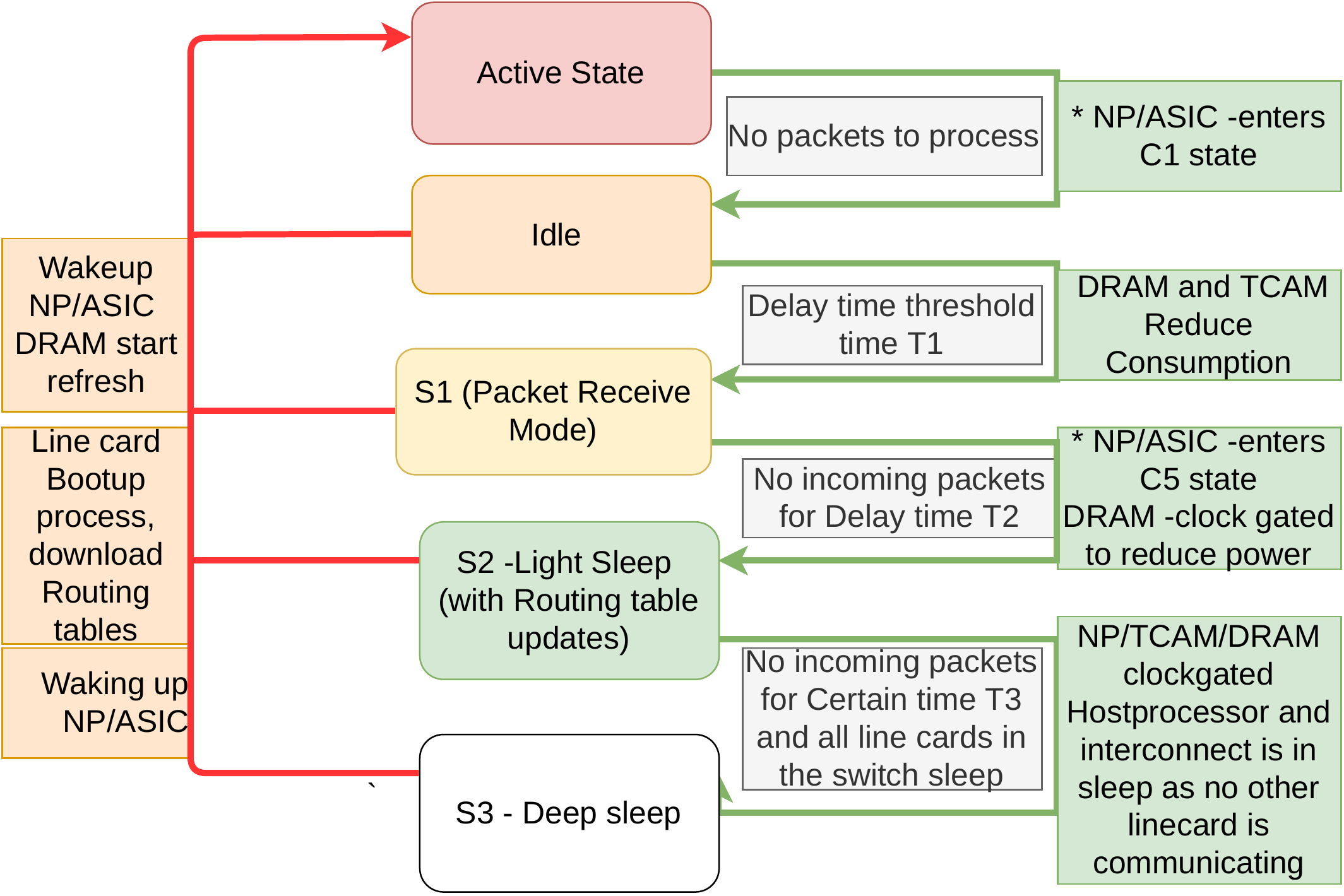}\caption{Illustration of a Switch Sleep and Wakeup Transition process with 3 sleep states.}
\label{figure13-sleepcycle}
\end{figure}

As shown in the Figure \ref{figure13-sleepcycle}, we model the sleep policy for a switch based on the power model described in section~\ref{sec:switchpower}. The line card unit of a switch is considered to be in active state when processing any output traffic, and idle when either there is no incoming traffic for certain period of time. In the first sleep state, the switch stops processing incoming packets but continues to receive new packets in the packet buffer to require the DRAM to be on the State. TCAM is not flushed to avoid the wakeup latency associated with re-populating the routing tables. In the second sleep state S2, the server's Energy consumption is reduced by setting the ASIC or Network processor to clock-gated state. In the second sleep state, additional savings can be achieved by turning off the DRAM, and deeper sleep state for the NP/ASIC. In the last sleep state, we clock gate the TCAM memories storing the routing tables, host processor and switch interconnects to be put in sleep state.

%% file: problemstatement.tex
\section{Design}
\subsection{Problem Definition}
\label{sec:problem solution}
In this section, we formulate the joint server and data center network power
optimization as a constrainted optimization problem using the switch power model
and job model defined above. First, to obtain the power consumption of a switch
$k$, assume the number of active line cards and ports are $\zeta^{active}_k$ and
$\rho^{active}_k$ respectively, and the number of line cards in respective sleep states and ports in LPI mode are $\zeta^{sleep}_k$ and $\rho^{LPI}_k$ respectively. Since the total power of a switch is the sum of base power $P_k^{base}$, power of ports, and power of line cards, then we have $P_k^{switch}=P_k^{base}+\zeta^{active}_k*P_{line card}^{active}+\rho^{active}_k*P_{port}^{active}+\zeta^{sleep}_k*P^{sleep}_{line card}+\rho^{LPI}_k*P_{port}^{LPI}$.
To calculate the power consumption of server $i$, since our system consideres the core as basic processing unit in server, the total power of a server is the sum of idle power $P_i^{idle}$ and dynamic power which is linear in the number of active cores $C_i^{on}$. Then we have $P_i^{server}=P_i^{idle}+C_i^{on}*P_{core}^{on}$, where $P_{core}^{on}$ denotes the power consumed by an active core. 

Then the joint power optimization problem can be formulated as minimize
$\sum_{k=1}^{N_{switch}}{P_k^{switch}}+\sum_{i=1}^{N_{server}}{P_i^{server}}$
under both network-side and serve-side constraints, such as link capacity,
computation resources, etc.

\begin{table*}
	\centering
	\scriptsize
	\caption{Estimating the energy consumption}
	\label{tab:estimatingenergy}
	\arrayrulecolor{black}
	\begin{tabular}{|l|l|l|} 
		\hline
		\multicolumn{2}{|c|}{\textbf{Energy Component} }                                                                                                                                                                                         & \textbf{Description}                                                                                                                                                                                        \\ 
		\hline
		\multicolumn{3}{|l|}{\textbf{~ ~ ~ ~ ~ ~ ~ ~ ~ ~ ~ ~ ~ ~ ~ ~ ~ ~ ~ ~ ~ ~ ~ ~ ~ ~ ~ ~ ~ ~ ~ ~ ~ ~ ~ ~ ~ Server Energy} }                                                                                                                                                                                                                                                                                                                                \\ 
		\hline
		\multirow{2}{*}{\textbf{Candidate server} }                                                                              & Server Task Execution Energy                                                                                  & Energy consumed when the task is being executed                                                                                                                                                             \\ 
		\arrayrulecolor[rgb]{0.651,0.651,0.651}\cline{2-3}
		& Server waiting for tranmission                                                                                & Energy consumed when the task is waiting for communication to complete.                                                                                                                                     \\ 
		\hline
		\multirow{3}{*}{\textbf{Server (If sleeping)} }                                                                          & Wakeup Energy during transistion                                                                              & Energy spent during wakeup.                                                                                                                                                                                 \\ 
		\cline{2-3}
		& Core Energy savings loss Sleep                                                                                & If sleeping, Energy savings loss due to Active state                                                                                                                                                        \\ 
		\cline{2-3}
		& Package Energy savings loss Sleep                                                                             & If the first core being woken up, CPU package sleep savings loss.                                                                                                                                           \\ 
		\arrayrulecolor{black}\hline
		\multicolumn{3}{|l|}{\textbf{~ ~ ~ ~ ~ ~ ~ ~ ~ ~ ~ ~ ~ ~ ~ ~ ~ ~ ~ ~ ~ ~ ~ ~ ~ ~ ~ ~ ~ ~ ~ ~ ~ ~ ~ ~Network Energy} }                                                                                                                                                                                                                                                                                                                                  \\ 
		\hline
		\multirow{3}{*}{\begin{tabular}[c]{@{}l@{}}\textbf{For Each Switch in the }\\\textbf{candidate flow path} \end{tabular}} & \begin{tabular}[c]{@{}l@{}}Network Energy cost based \\on Bandwidth-allocated \end{tabular}                   & \begin{tabular}[c]{@{}l@{}}Every flow is allocated a BW based on the network congestion and~\\link capacity of the entire flow path,~flow's minimum bandwidth \\the requirement to meet QoS. \end{tabular}  \\ 
		\arrayrulecolor[rgb]{0.651,0.651,0.651}\cline{2-3}
		& Wakeup Energy during transistion                                                                              & Energy spent to transistion to active state                                                                                                                                                                 \\ 
		\cline{2-3}
		& Line card Active energy                                                                                       & If the Linecard is sleeping, the energy cost is the full active power consumption.                                                                                                                          \\ 
		\hline
		\begin{tabular}[c]{@{}l@{}}\textbf{For Every flow in common }\\\textbf{path of the candidate path} \end{tabular}         & \begin{tabular}[c]{@{}l@{}}Energy savings due to decreased \\Bandwidth for the every other flow \end{tabular} & \begin{tabular}[c]{@{}l@{}}For every other switch in the current flow path whose BW is reduced,\\calculate the energy spent on it. \end{tabular}                                                            \\
		\arrayrulecolor{black}\hline
	\end{tabular}
\end{table*}

%% file: algorithms.tex
\subsection{Estimating the energy consumption}
\label{sec:algorithms}
Modeling this joint power optimization problem in Datacenter Networks (DCN) as an Integer Linear Programming (ILP) formulation is a solution, and optimization tools like MathProg can be used to provide a near-optimal result. However, the computation complexity increases exponentially with the number of servers and switches~\cite{wangExpeditusCongestionAwareLoad2017}. In a typical data center with tens of thousands of servers and hundreds of switches, it is computationally prohibitive to solve the optimization problem. We therefore propose a computationally-efficient heuristic algorithm in this section.

\input{motivation}
\subsection{Heuristic Algorithms}
In this section, we first present a power management algorithm for the transition of line card and port power state, a simple power transition algorithm for servers, and then propose joint job placement and network routing algorithm for solving the optimization problem efficiently.


\subsubsection{Power State Transition Algorithm}


As not all switches in DCN need to be active all the time, if we can intelligently control the transition of active and low power state for ports and line cards, then data center network power consumption could be saved. Therefore, we propose the Power State Transition Algorithm~\ref{algorithm1} to implement network power management. The notations are elaborated in Table~\ref{algorithm1table}. We assume that there is a global controller that keeps record of all the line cards, ports, and server status, including their power state, queue size, etc. The global controller monitors the current traffic load (number of pending flows or packets) at each port and decides whether the current line card power state should change. It then decides if the port state should change as a feedback. In our design, we couple the line card and port power state and their transition: if a line card is in sleep or off state, then all the ports are also in LPI or off state; if a line card is active, then its ports can be in LPI or active state, and LPI ports can be woken up to become active in a short time. 


\begin{algorithm}
	\SetAlgoNoLine
	\scriptsize
	\caption{Power State Transition Algorithm}
	\label{algorithm1}
	\KwIn{$T_s$, $T_a$, $Q_{iL}$}
	\KwOut{Line card and port state transition}
	Initialization: All line cards are in sleep state, all ports are in LPI state\;
	
	\While{there are jobs to be executed}
	{
		\If{a flow arrives at port $i$ of line card $L$ at time $t$}
		{
			\If{$L$ is in sleep state} {
				\If{$Q_{iL} > T_s$} {
					$L$ begins waking up from sleep state\;
					$i$ begins waking up from LPI state\;
				} 
				\Else
				{
					$L$ begins waking up after $\tau_{wakeup}^{LC}$\;
					$i$ begins waking up after $\tau_{wakeup}^{port}$\;
				}
				
			} 
		}
		\If{a flow is transimitted from port $i$ of line card $L$ at time $t$} {
			\If{$Q_{iL} < T_s$} {
				$i$ starts transition to LPI after $\tau_{LPI}$\;
				\If{after $\tau_{LPI}$, all the ports of $L$ are in LPI state} {
					$L$ starts transition to deeper sleep state after $\tau_{sleep}^{LC}$;
				}
			}
		}
	}
	
\end{algorithm}

%
%
%
%
%

\subsubsection{\PopCorns-Cooperative Network-Server Algorithm}

{

The main idea of our algorithm is to jointly consider the status of server pool and network before assigning jobs. To be more specific, for a job consisting of several pairs of interdependent tasks, if we place task pairs based on their interdependence, and choose the core pair with the minimum routing cost, instead of randomly placing them on available cores without awareness of communication requirement between tasks, then the placement together with its corresponding routing path must be the optimal. And in the context of this paper, network routing cost is energy consumption, since every line card on the chosen routing path has to be active, or it will have to be woken up.

Based on this idea, we propose the Cooperative Network-Server (CNS) Algorithm~\ref{algorithm2} 
The notations are elaborated in Table~\ref{algorithm1table}. In the initial stage, we compute and store all possible routing paths between any pair of network nodes. As mentioned in the previous section, we have a global controller to keep track of all the line cards, ports, and server statuses. Thus, when a job consisting of a set of interdependent tasks arrive, we first check the server side to select all server pairs whose power state is active and local queue sizes do not exceed a threshold. If no server statisfies these requirements, then servers with full local queue and servers in C6 sleep state will be selected. Note that if a task is assigned to a server in sleep state, then the server will be activated immediately and enter active state after a wakeup latency. For each server pair we select possible routing paths between them from the pre-computed routing paths set. Along each path, line cards could be active, sleeping, or off, and ports could be active, LPI, or off, but if we assign a path to the task pair, all the line cards and ports along it should become active. In other words, we need to wake up the inactive line cards and ports on the chosen path, which results in extra power consumption and wakeup latency. Based on this, we can compute and assign a weight, which is a measurement of the number of active line cards and ports, to each possible routing path, and choose the server pair with minimum routing cost. And our proposed Cooperative Network-Server Algorithm will output the least-weight path and its corresponding server pair for each task pair.}

\begin{table}
	\centering
	\captionsetup{font=small}
		\scalebox{0.7}{
	\begin{tabular}{|c|c|c|}
		\hline
		\bf{Symbol}& \bf{Description} \\ 
		\hline
		$P$ & all the routing paths between any pair of network nodes  \\
		\hline
		$Q_s$ & local queue size of server $s$  \\
		\hline
		$T_s^{server}$ & local queue size threshold of server $s$\\
		\hline
		$S$ & all the servers in DCN  \\
		\hline
		$S_{ava}$ & all the servers whose current queue size doesn't exceed $T_s^{server}$  \\
		\hline
		$P_{x, y}$ & all the routing paths between node $x$ and $y$ in DCN\\
		\hline
			$T_s$ & traffic threshold for port waking up  \\
		\hline
		$T_a$ & traffic threshold for port turning into LPI state  \\
		\hline
		$Q_{iL}$& current traffic load at port $i$ in line card $L$  \\
		\hline
		$\tau_{wakeup}^{port}$ & port wakeup delay  \\
		\hline
		$\tau_{wakeup}^{LC}$ & line card wakeup delay  \\
		\hline
		$\tau_{LPI}^{port}$ & port turning into LPI state delay \\
		\hline
		$\tau_{sleep}^{LC}$ & line card turning into sleep state delay \\
		\hline	
	\end{tabular}
}
	\caption{\label{algorithm1table} Notations in \PopCorns Cooperative Network-Server Algorithm.}
\end{table}

\begin{algorithm}[!htb]
	\SetAlgoNoLine
	\scriptsize
	\caption{\PopCorns  - Cooperative Network-Server Algorithm}
	\label{algorithm2}
	\KwIn{$P$, $Q_s$, $T_s$, line cards and ports power state, task dependency within a job}
	\KwOut{Job placement and corresponding routing path}
	
	\While{job $j$ consisting of task set $T^j$ arrives}
	{
		\For{each pair of interdependent tasks $(T^j_m, T^j_n)$ in $T^j$}
		{
			select $S_{ava}$ from $S$\;
			\For {each pair of available servers $(x, y)$ in $S_{ava}$}
			{
				compute $P(x, y)$ for server pair $(x, y)$ from $P$\;
				\For {each path $p$ in $P(x, y)$}
				{
					get power states of all the ports and line cards along $P$ from the global controller, compute path weight $w(x, y)$ in terms of energy consumption\;
				}
			}
			choose the least-weight path associated with corresponding server pair for task pair $(T^j_m, T^j_n)$\;
		}
	}
\end{algorithm}

\begin{algorithm}
	\SetAlgoNoLine
	\scriptsize
	\caption{\PopCorns - Weight assignment algorithm between two nodes}
	\label{algorithm3}
	\KwIn{$Link(i,j)$, $QoS$, $Link_{Capacity}$}
	\KwOut{$W(i,j)$ Weight for the edge connecting nodes i and j}

	\If{ Node $i$ is Server} {
		 $LinkWeight(i,j) += (ServerActivePower - CurrentSleepStatePower) *( taskSize + EdgeLinkBW / FlowSize)$;
	 }
	\If{ Node $j$ is a Switch} 
	{
		$LinkCapacity_{remaining} \leftarrow Link_{fullCapacity}$;
		
		\For {$Every Flow F_{i}  on  Link(i,j)$ }
			{
				$Flow_{i_{remainingTime}} \leftarrow  FSize_{i_{Remaining}} / FcurBW_{i}$;

				\If {$F_{i_{remainingTime}} > SlowestTime$ } 
					{
					 $SlowestTime \leftarrow  F_{i_{remainingTime}}$;
					 
					 $SlowestFlow \leftarrow F_{i}$;
					 
					}
				$LinkCapacity_{remaining} -= FcurBW_{i}$
			}
		$availBW \leftarrow min(Link_{fullCapacity} - LinkCapacity_{remaining} \text{ and } EdgeLinkBW)$;
		
		\If { $availBW > MinBWforQoS$ }
		{ 
			$timeForNewFlow \leftarrow availBW/ FSize_{newFlow}$;
			
			$additionalTimeAwake \leftarrow MAX(SlowestTime \text{ and } timeForNewFlow  )$;
			
			$LinkWeight(i,j) += additionalTimeAwake * ActivePower_{j}$;
		} \Else {
			$FcurBW_{newFlow} \leftarrow FSize_{newFlow}/QoSRequiredTime$;
			
			$FcurBW_{Slowest}' \leftarrow (FcurBW_{Slowest} - (availBW - (FcurBW_{newFLow}/NumFlowsOnLink))$;
		
			$AdditionalTime_{SlowestFlow} \leftarrow       (FcurBW_{Slowest} - FcurBW_{Slowest}')/ FlowDataremaing_{Slowest}$;
			
			$timeForNewFlow \leftarrow QoSRequiredTime$
			
			$additionalTimeAwake \leftarrow MAX(timeForNewFlow \text{ and } AdditionalTime_{SlowestFlow})$;
			
			$LinkWeight(i,j) += additionalTimeAwake * ActivePower_{j}$;
		 
		}
	}
\end{algorithm}

\begin{algorithm}
	\SetAlgoNoLine
	\caption{Job scheduling Algorithm}
	\label{algorithm2}
	\scriptsize
	\KwIn{$P$, $Q_s$, $T_s$, line cards and ports power state, task dependency within a job}
	\KwOut{Job placement and corresponding routing path}

	\While{job $j$ consisting of task set $T^j$ arrives}
	{
		\For{each pair of interdependent tasks $(T^j_m, T^j_n)$ in $T^j$}
		{
			select $S_{ava}$ from $S$\;
			\For {each pair of available servers $(x, y)$ in $S_{ava}$}
			{
				compute $P(x, y)$ for server pair $(x, y)$ from $P$\;
				\For {each path $p$ in $P(x, y)$}
				{
					get power states of all the ports and line cards along $P$ from the global controller, compute path weight $w(x, y)$ in terms of energy consumption\;
				}
			}
			choose the least-weight path associated with corresponding server pair for task pair $(T^j_m, T^j_n)$\;
		}
	}
\end{algorithm}

\todo{TBD}


%% file: motivation.tex
%

\subsection{Need for Switch sleep states}
\label{sec:motivation}
In this section, we motivate the need for having more sleep states in switches. As discussed in sections \ref{sec:transition} and \ref{sec:switchpower}, the switch power consumption can be temporarily reduced by selectively turning off parts of the line cards and woken up when required. For components which do not have any memory or state the wakeup latency is the time it takes to initialize the system. For other components such as DRAM storing address forwarding tables should be reinitialized from the host line cards. In our study, we find that having multiple sleep states for the switch helps in trade-off various levels of wake-up latency for energy consumption reduction. This scheme is inherently useful when the idle period is not large enough to accommodate the high wastage of energy spent during waking from a single deepest sleep state and would cause network transmission to get delayed. 

\begin{figure}[!h]
	\captionsetup{font=small}
	\centering\includegraphics[width=0.4\textwidth]{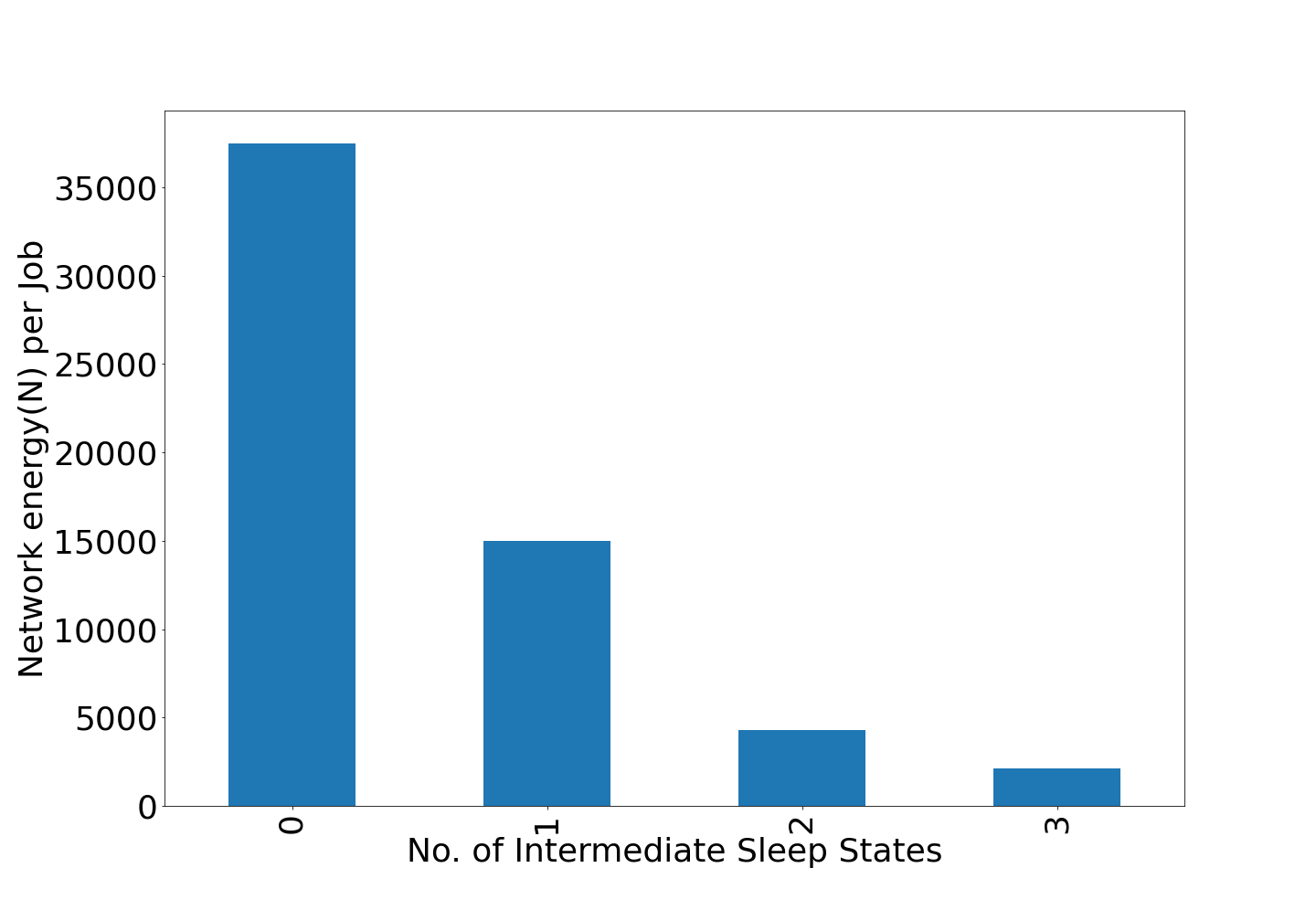}\caption{Network Energy consumption normalized per job for a 1024 server FatTree topology network. Jobs containing two dependent tasks of 500ms CPU duration each arrive at 30 Job/second. The Switch sleep state and latency models are constructed using the power model defined in section~\ref{sec:switchpower}.} 
	\label{motivation12}
\end{figure}

%% file: experiment.tex
\section{Evaluation}
\label{sec:experiment}
\subsection{Experimental Setup}
Our experiments were performed using HolDCSim, an event-driven simulator~\cite{HolDCSimHolisticSimulator},
which simulates interdependent asynchronous task executions, and implement
custom server job scheduling and network routing algorithms. This allows us to
simulate a large number of servers and switches relatively easily. The simulator
allows us to specify our own power model and sleep state transition mechanisms
and calculate the overall energy consumption for the simulator. The network is
configured using a fat tree-topology as shown in Figure~\ref{figure12}.  We
simulate two classes of real-world applications: 
\emph{web service}: two large sized (uniformly distributed around 500 ms service time) parent-child tasks,
\emph{web search}: a parent task search task querying five subtasks to run smaller search process (uniformly distributed around 100 ms CPU service time). All the network routing policies consider the QoS threshold of 10X the total CPU time of all tasks in the job when choosing the job selection.

\begin{figure} [!h]
	\centering
	\captionsetup{font=small}
	\subfloat[Steps in a web search request.]{		\includegraphics[scale=0.34]{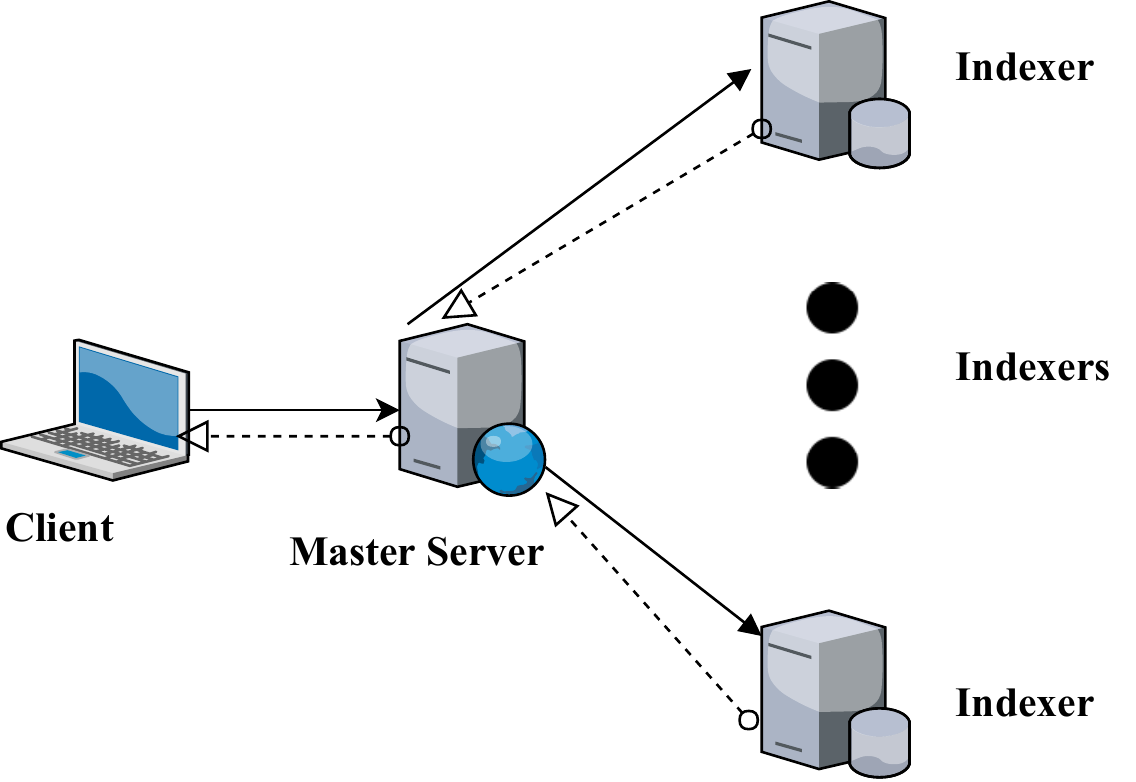}
		\label{fig:ws-figure90}
}
	\subfloat[DAG of a web search job.]{		\includegraphics[scale=0.15]{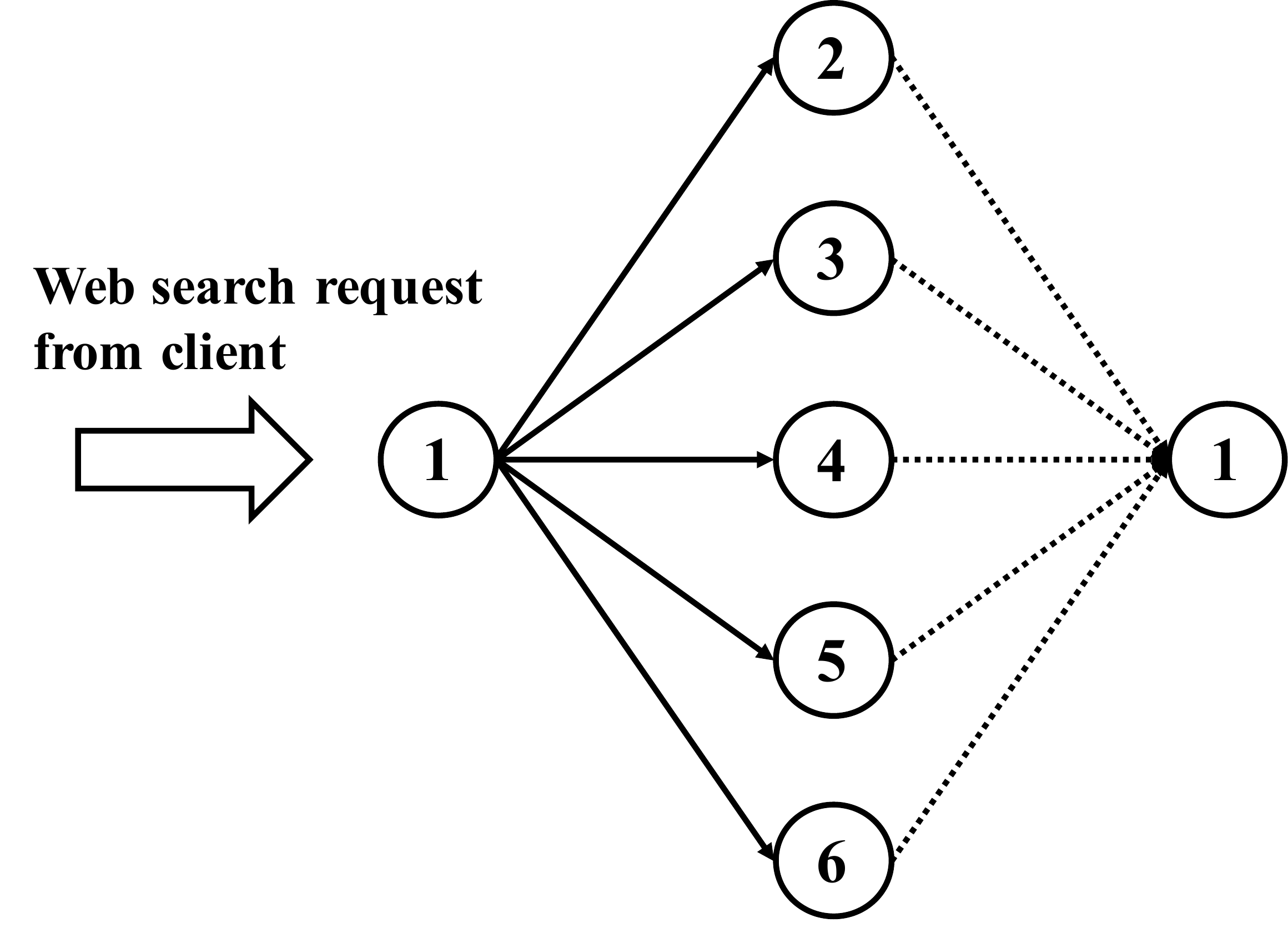}
		\label{fig:ws-dag}
}
	\caption{Web search communication patterns (left) and DAG of a web search job (right). 
		Task 1 is executed by the master server, while task 2 - 6 are processed by different indexers, and master server communicates with all the indices.}
	\label{fig:ws-graph}
\end{figure}

\begin{figure} [!h]
	\centering
	\captionsetup{font=small}
	\subfloat[Steps in a web service request.]{
		\includegraphics[scale=0.34]{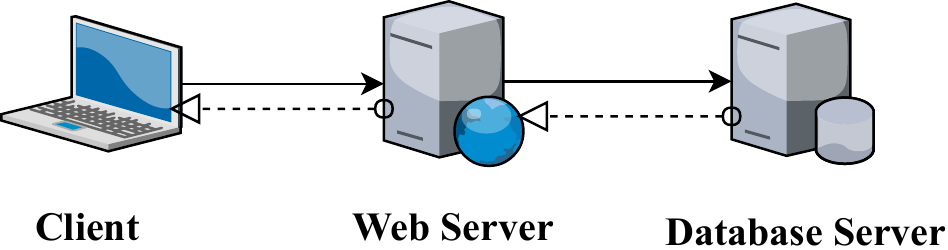}
		\label{fig:wsv-figure90}
s	}
	\subfloat[DAG of a web service job.]{
		\includegraphics[scale=0.15]{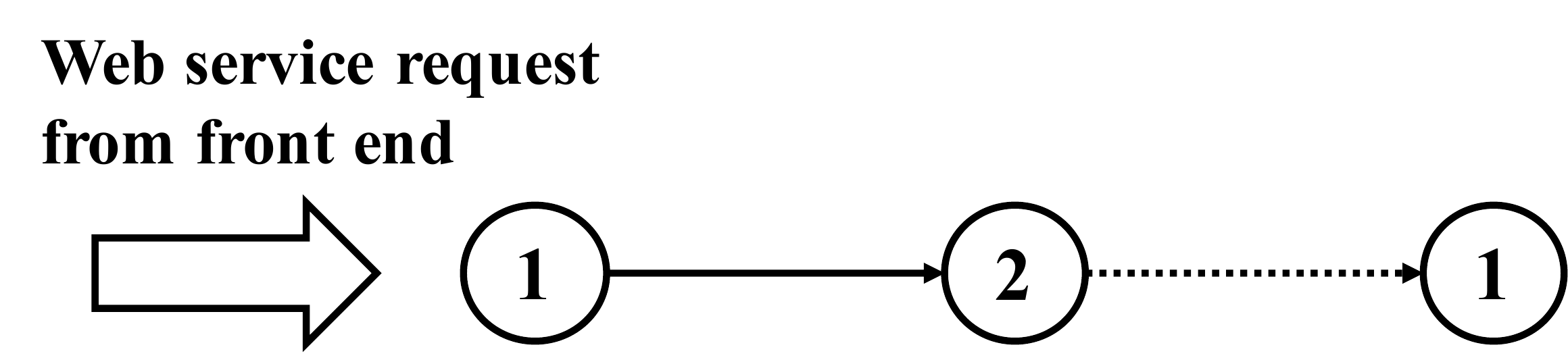}
		\label{fig:wsv-dag}
	}
	\caption{Web service communication patterns (top) and DAG of a web service job (bottom). 
		Task 1 is executed by application server, while task 2 is assigned to database server.}
	\label{fig:wsv-graph}
\end{figure}

We assume that each of the tasks can be executed on any server machine and if enough processing cores available co-located in the same server.  The communication patterns and the DAG for the two workloads are illustrated in Figure~\ref{fig:ws-graph}, and Figure~\ref{fig:wsv-graph}.

We carefully select delay timer values for sleep state transitions~\cite{yaoWASPWorkloadAdaptive2017} for server and network by trying each value in the search-space, starting with first determining the delay timer value for Sleep state 1 with the lowest energy consumption, then fixing the lowest timer value found previously for state 1 to determine the delay timer for sleep state 2, and so on.


\subsubsection{Job arrivals patterns}
\label{workloads}

We use two kinds of job arrivals in our study.
The arrival rate is given by $\lambda = (Num_Servers * Num * Cores * \rho)/ Avg_TaskSize * NumTasksperJob$. Thereby targeting $\rho$ values of 15\%, 30\% and 60\% we derive the $\lambda$ values. Due to the variation in the scheduling policy, task inter-dependency, randomized job task and arrival times and sleep state transitions the resultant server utilization level may not match with targeted utilization levels.

\emph{Markov Modulated Poisson Process}: Similar to workload used in \cite{yaoWASPWorkloadAdaptive2017}, MMPP arrivals generates a two different states of arrival rates. The normal utilization phase similar to the Poisson workload and a bursty period where the arrival rate is 1.5X the $\lambda$ in the normal phase. We choose the bursty period to last 10 secs between 20 seconds of normal phase. The load levels of low, medium and high arrival rate and obtained in the similar fashion as the basic poisson based arrival workload.

 \emph{Trace based arrival job arrival pattern}: We use publicly available job arrival trace from NLANR~\cite{NationalLabApplied} for a real world system. We chose NLANR trace since it inherently has a high variation in job inter-arrival rate. The load levels of low, medium and high arrival traces are obtained by scaling the inter-arrival times between the jobs in the original trace by a factor.


\subsubsection{Quality of Service Constraints}
We consider the minimum time a job takes to complete to be the sum of computation time of each task in the job and time it takes for all the inter-task communications to finish at the fastest possible bandwidth in the network. For instance, for a two-task job with 500ms of CPU cost each and a 10 MB of network communication on the 10 Gbps maximum link capacity between two nodes, the minimum time for job execution will be \textit{500ms x 2 + 75 ms (data transfer time) = 1075 ms}. Similar to other works (e.g.,~\cite{yaoWASPWorkloadAdaptive2017}), we consider the quality of Service (QoS) deadline for the job scheduler to be 10 X the minimum job execution to enable enough slack to allow multiple jobs and enable energy optimizations.

\subsection{Baseline policies and Evaluation methodology}

There are two major objectives: first to prove that the smart use of line card sleep states with our proposed Algorithm~\ref{algorithm1}, does reduce power consumption in Popcorns-Pro compared to no power management policy on the switches at all; second, our proposed \PopCorns 
algorithm~\ref{algorithm2} can further save power, compared to other job placement algorithms not taking both the network and server status into consideration. We consider the four combinations of the below server-specific and network-specific energy optimization policies. As shown in the section ~\ref{sec:motivation}, we have seen that switch sleep states without any optimization can still yield some energy savings, and it is important to further optimize the energy savings for them.

\subsubsection{Server based policies}
	The below two policies represent a common server selection policy and energy optimized server policy.
{\textbf{Random Server allocation}-\textit{SB}} 
Following a traditional server load balancing approach, tasks are assigned randomly to all servers and the processing cores as they come in, without any consideration for task dependency and core locality in the data centers. 

{\textbf{Workload Adaptive using Server Low-Power State Partitioning-\textit{WASP}}} This policy represents recent works~\cite{yaoWASPWorkloadAdaptive2017} which consolidate jobs into fewer server while optimizing for keeping more servers in sleeping state. These optimizations are typically developed by system developers without considering the network energy optimization. In WASP, a set of servers are kept in the unused state to allow them to achieve the most energy efficient sleep state, while the remaining jobs are processed on the remaining servers. The servers are reassigned between the two sets depending the measured input load on the entire server pool.

\subsubsection{Network based policies}
The following two approaches to network routing in the data center.

{\textbf{Shortest Path-\textit{SP}}} This baseline network path allocation uses Djikstra's algorithm to forward flows across the two end points. This policy represents the traditional network routing algorithm available which chooses the shortest number of hops which also considering the QoS latency requirements for the current flow. 

{\textbf{Elastic tree-\textit{ET}}} Recent work on the improving network energy efficiency have considered various heuristics to consolidate flows into fewer switches. The Elastictree ~\cite{hellerElasticTreeSavingEnergy2010} baseline represents one of the most commonly cited work in this area, which tries keep fewer switches in active state in highly redundant network typologies such as the fat tree topology. We implement the greedy bin packing approach in our experiment where the heuristic chooses the leftmost path at every switch exit to reach the destination. These approaches assume a fixed server task allocation and hence cannot fully optimize for efficiency.

\input{baselines}
service and its corresponding job DAG. In simulation, we set the flow size to be
100KBytes and average job service time is randomly generated between 150ms and 200ms based on prior studies~\cite{meisnerPowerNapEliminatingServer}.

\begin{figure*}[!h]
	\centering
	\captionsetup{font=small}
	\subfloat[Low Arrival Rate \newline $\lambda$= 8 Jobs/Sec]{
		\includegraphics[width=0.33\textwidth]{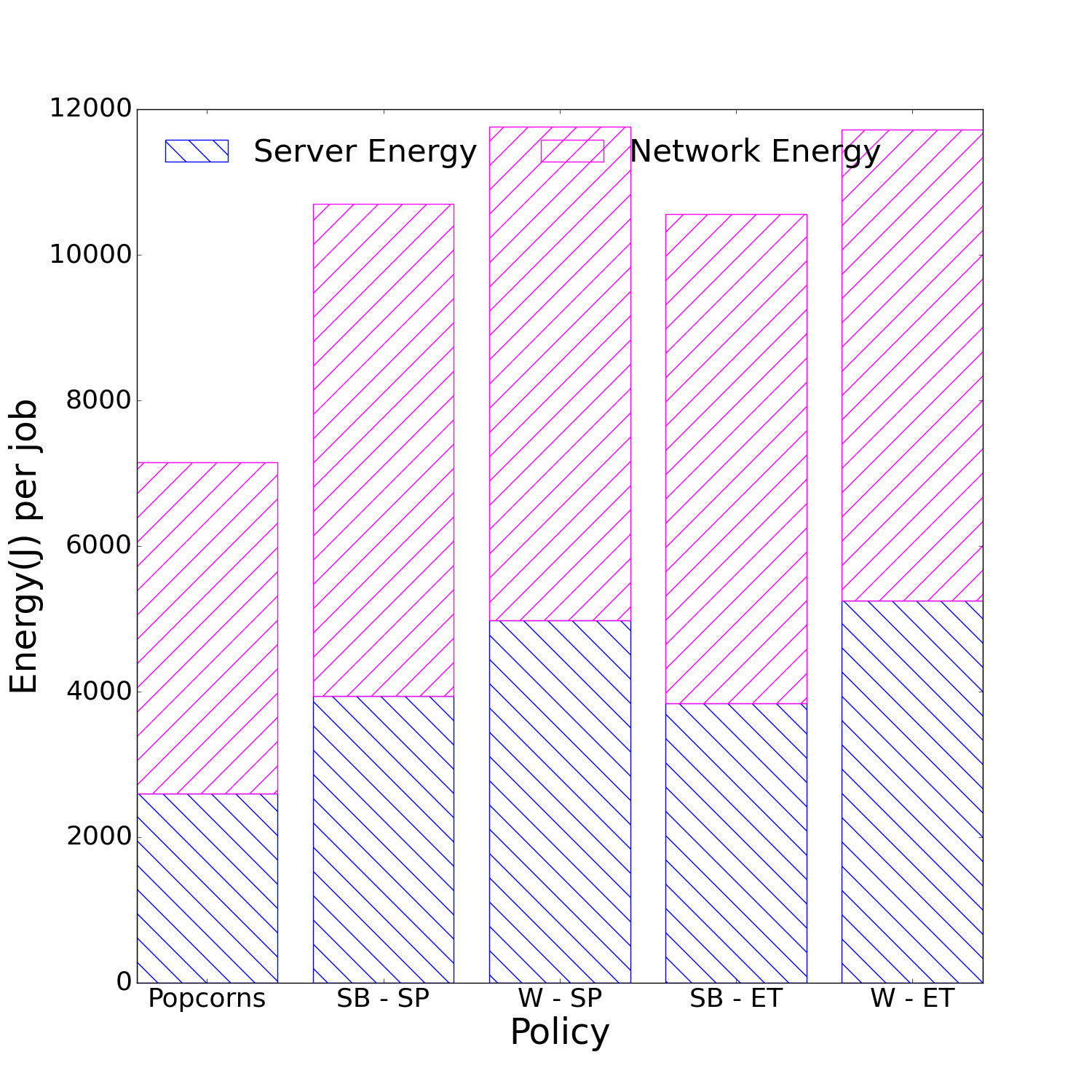}
		\label{djikstra_h_random}
	}
	 \subfloat[Medium Arrival Rate \newline $\lambda$ = 15 Jobs/Sec]{
	 	\includegraphics[width=0.33\textwidth]{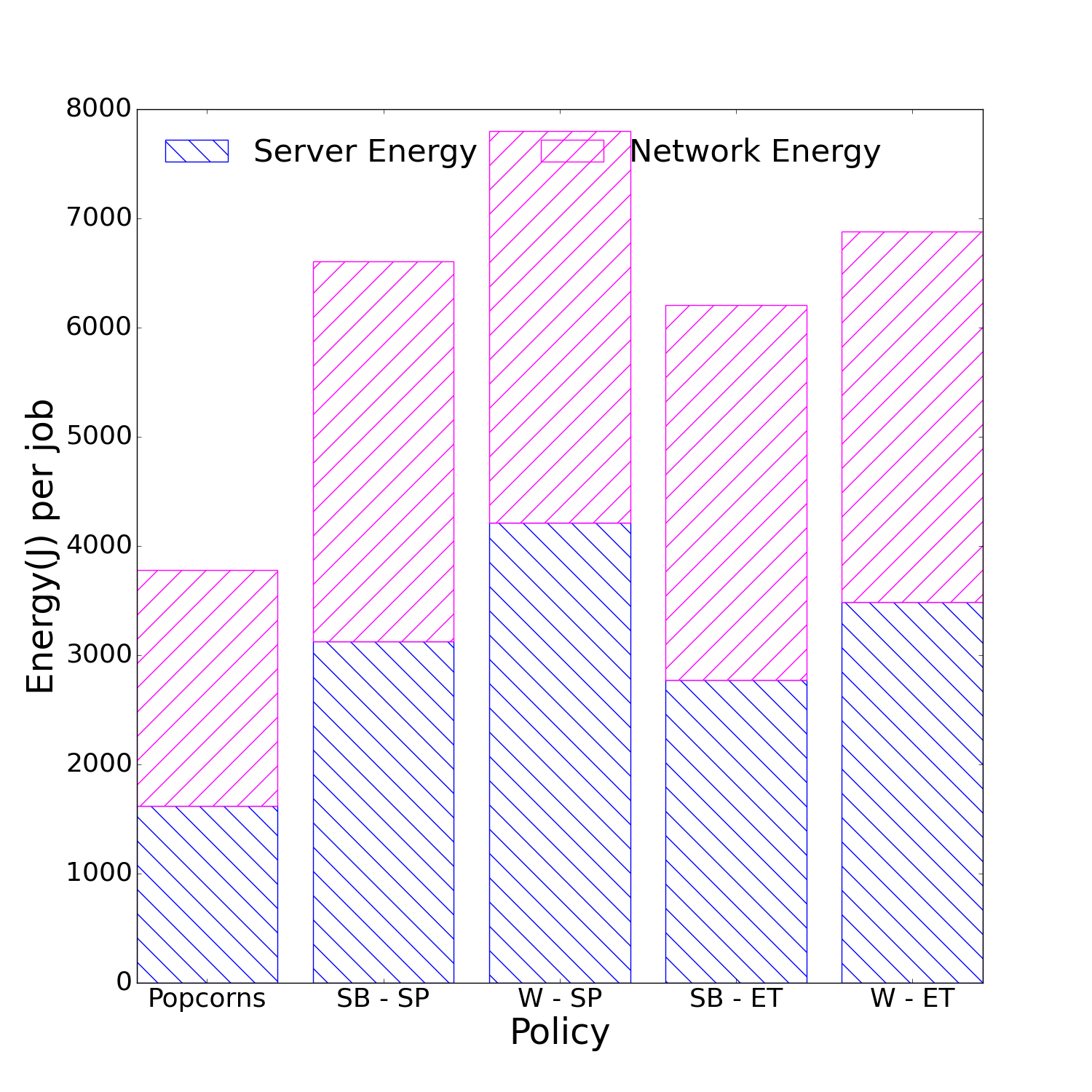}
	 	\label{elastic_tree_h_random}
	 }
	\subfloat[High Arrival rate \newline $\lambda$ = 30 Jobs/Sec ]{
		\includegraphics[width=0.33\textwidth]{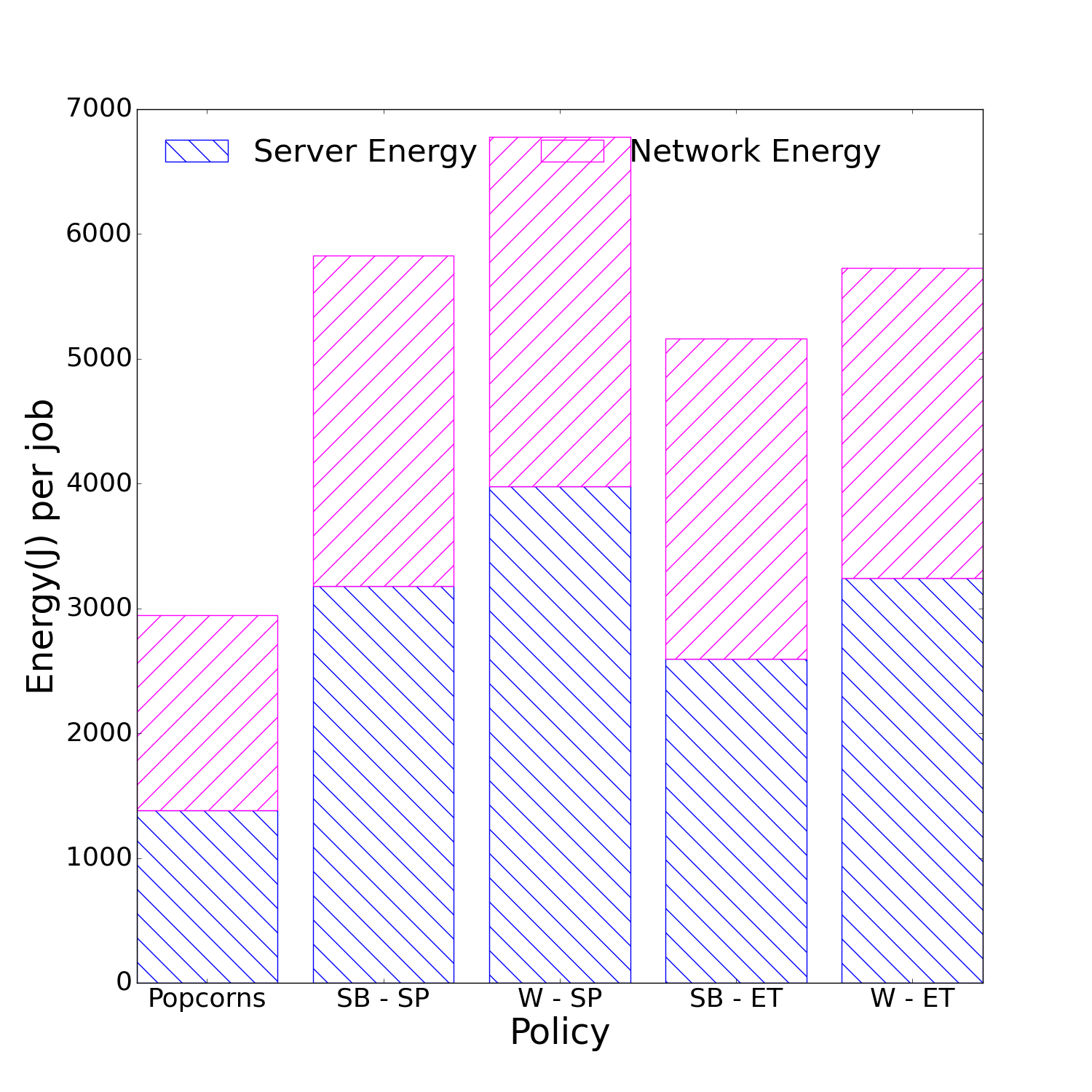}
		\label{elastic_tree_h_wasp}
	}
	\caption{Energy Savings comparison for different server and network path selection algorithms with MMPP based Job Arrival Pattern for service based 2 task model. Every 30 seconds we have 10-second bursts with 1.5x Arrival rate.}
	\label{fig:mmpp-ser}	
\end{figure*}
\begin{figure*}[!h]
	\centering
	\captionsetup{font=small}
	\subfloat[Low Arrival Rate \newline Avg. Rate = 10 Jobs/Sec ]{
		\includegraphics[width=0.33\textwidth]{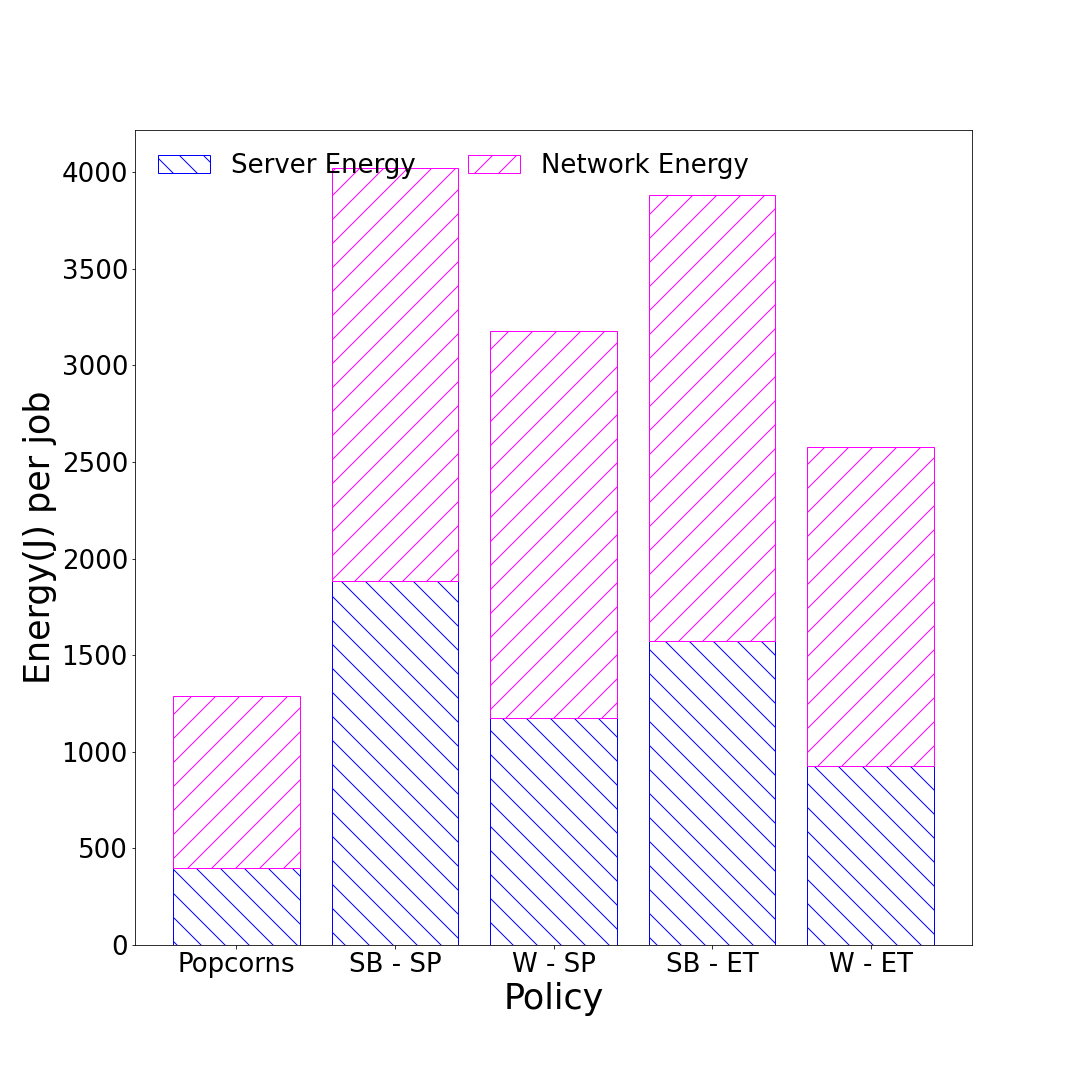}
		\label{djikstra_h_random}
	}
	 \subfloat[Medium Arrival Rate \newline Avg. Rate = 16 Jobs/Sec]{
	 	\includegraphics[width=0.33\textwidth]{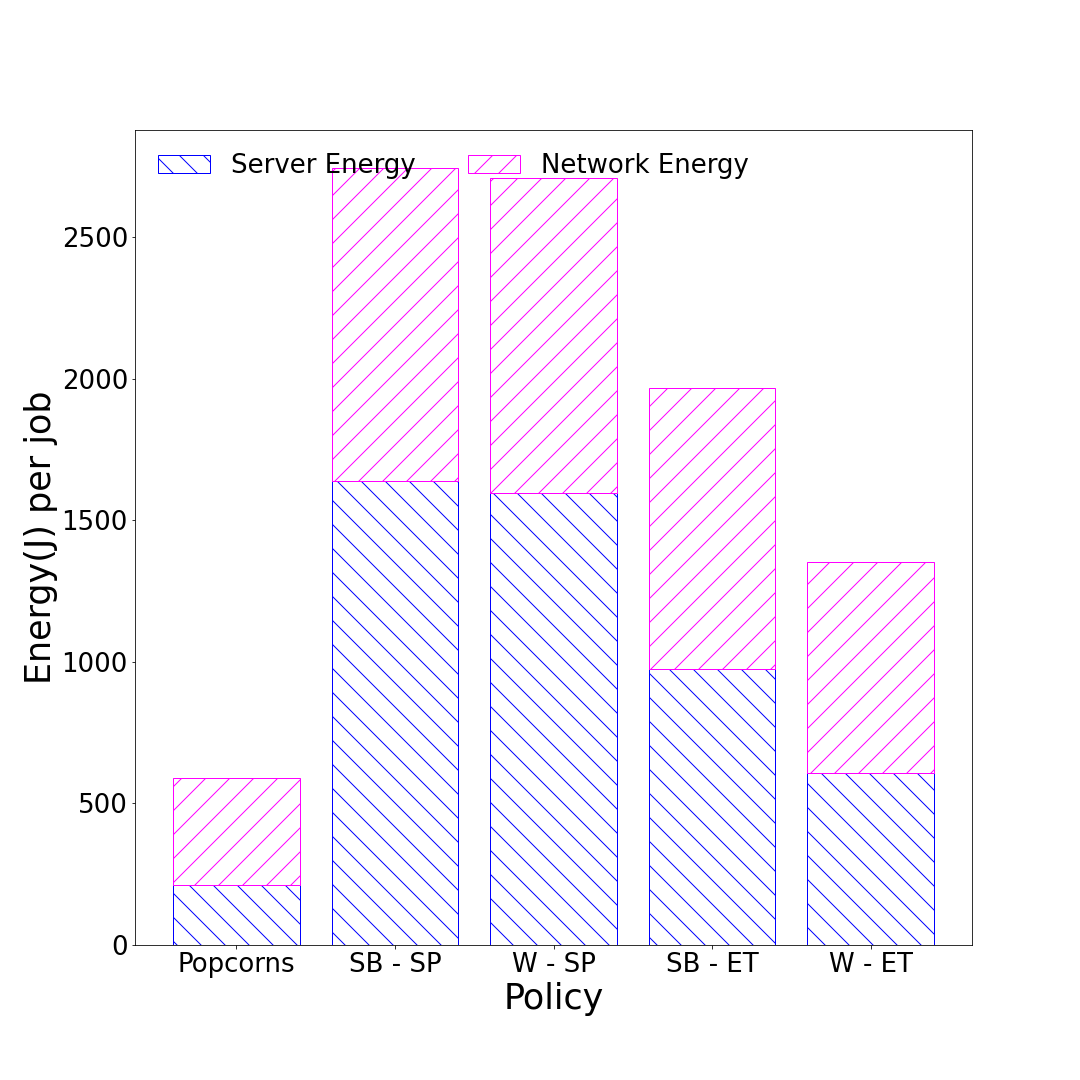}
	 	\label{elastic_tree_h_random}
	 }
	\subfloat[High Arrival rate \newline Avg. Rate= 32 Jobs/Sec]{
		\includegraphics[width=0.33\textwidth]{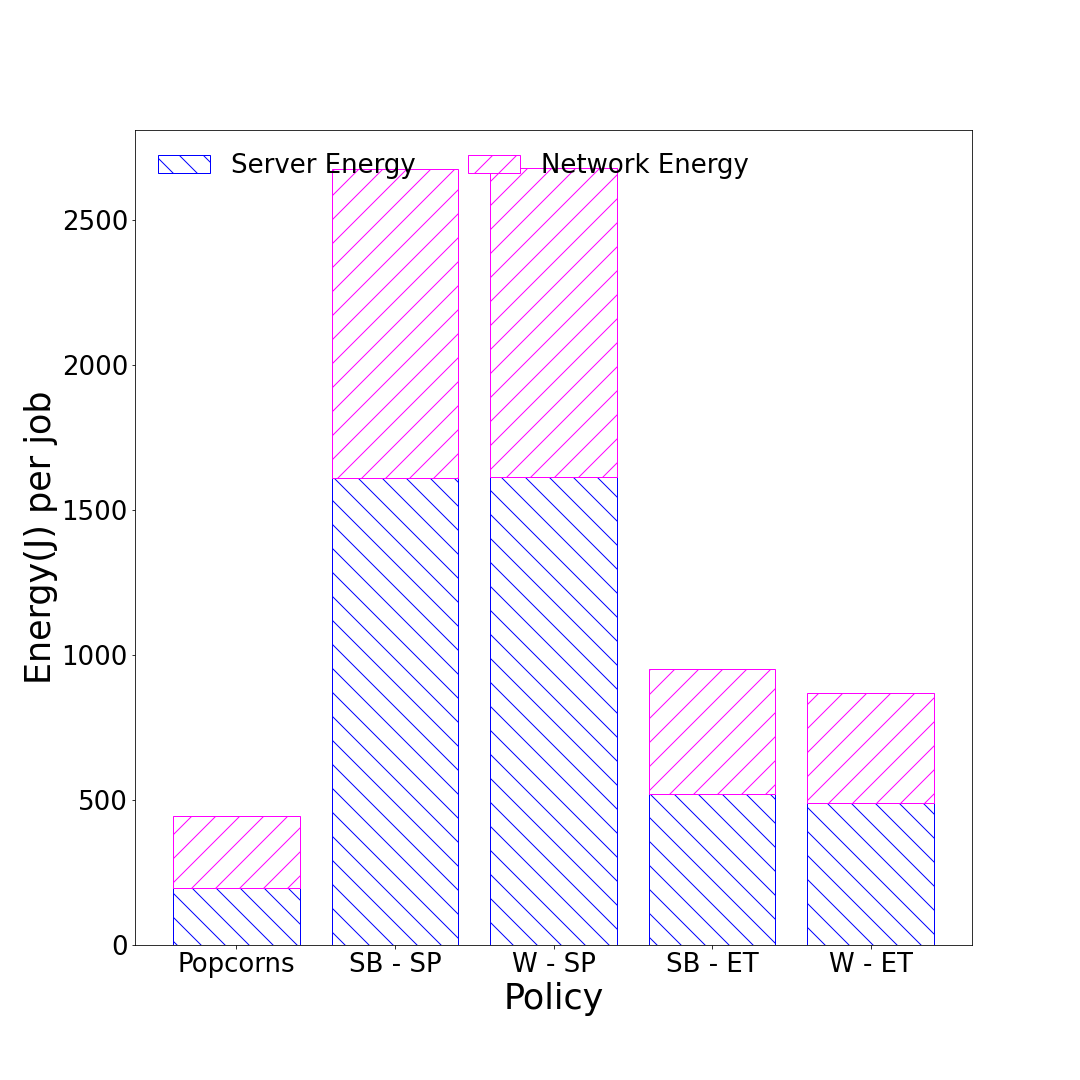}
		\label{elastic_tree_h_wasp}
	}
	
	\caption{Energy Savings comparison for different server and network path selection algorithms with NLANR trace based Job Arrival Pattern for service based 2 task model.} 
	\label{fig:trace-ser}
\end{figure*}

%
%

\begin{figure*}[!h]
	\centering
	\captionsetup{font=small}
	\subfloat[Low Arrival Rate \newline $\lambda$=8 Jobs/Sec]{
		\includegraphics[width=0.33\textwidth]{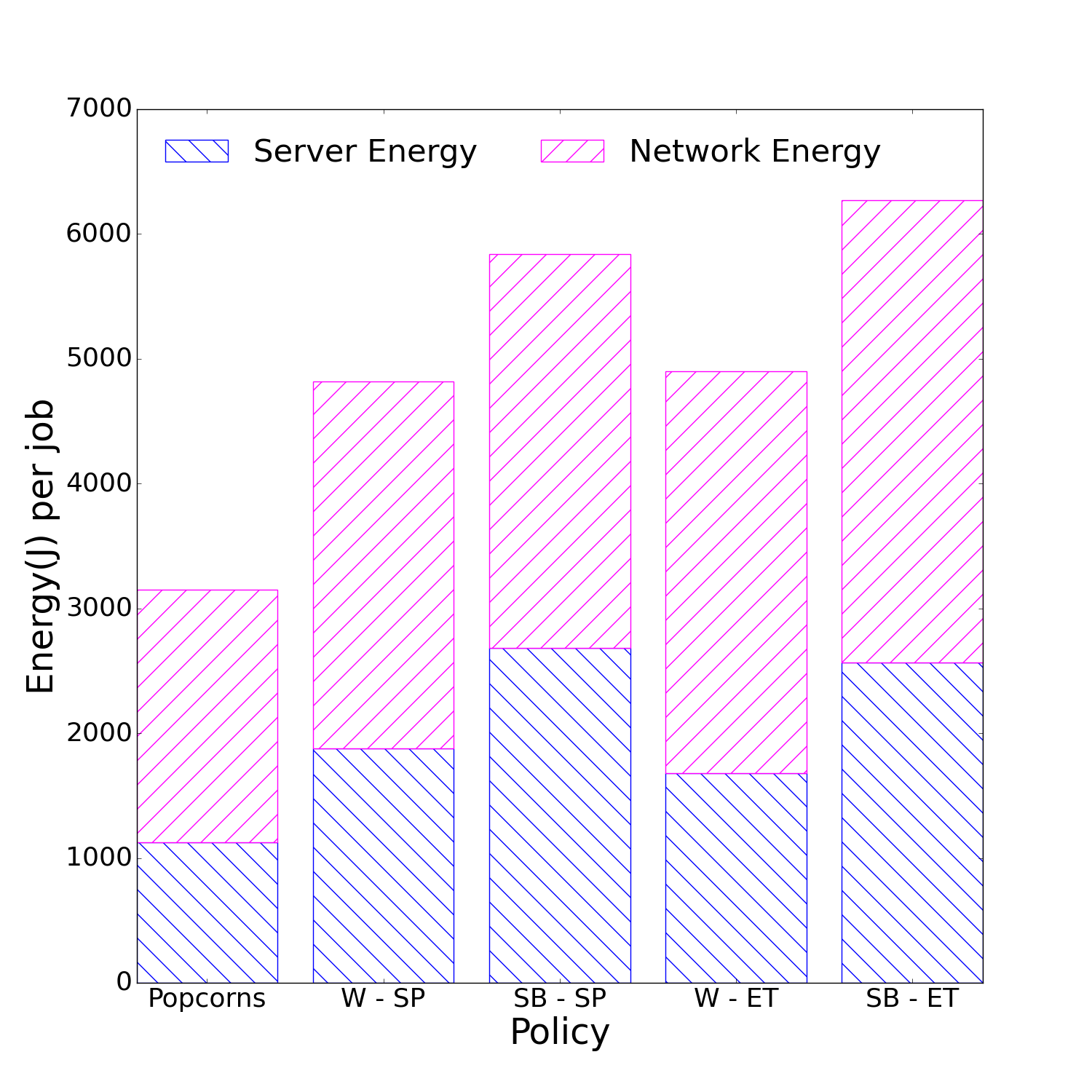}
		\label{djikstra_h_random}
	}
	 \subfloat[Medium Arrival Rate \newline $\lambda$=15 Jobs/Sec ]{
	 	\includegraphics[width=0.33\textwidth]{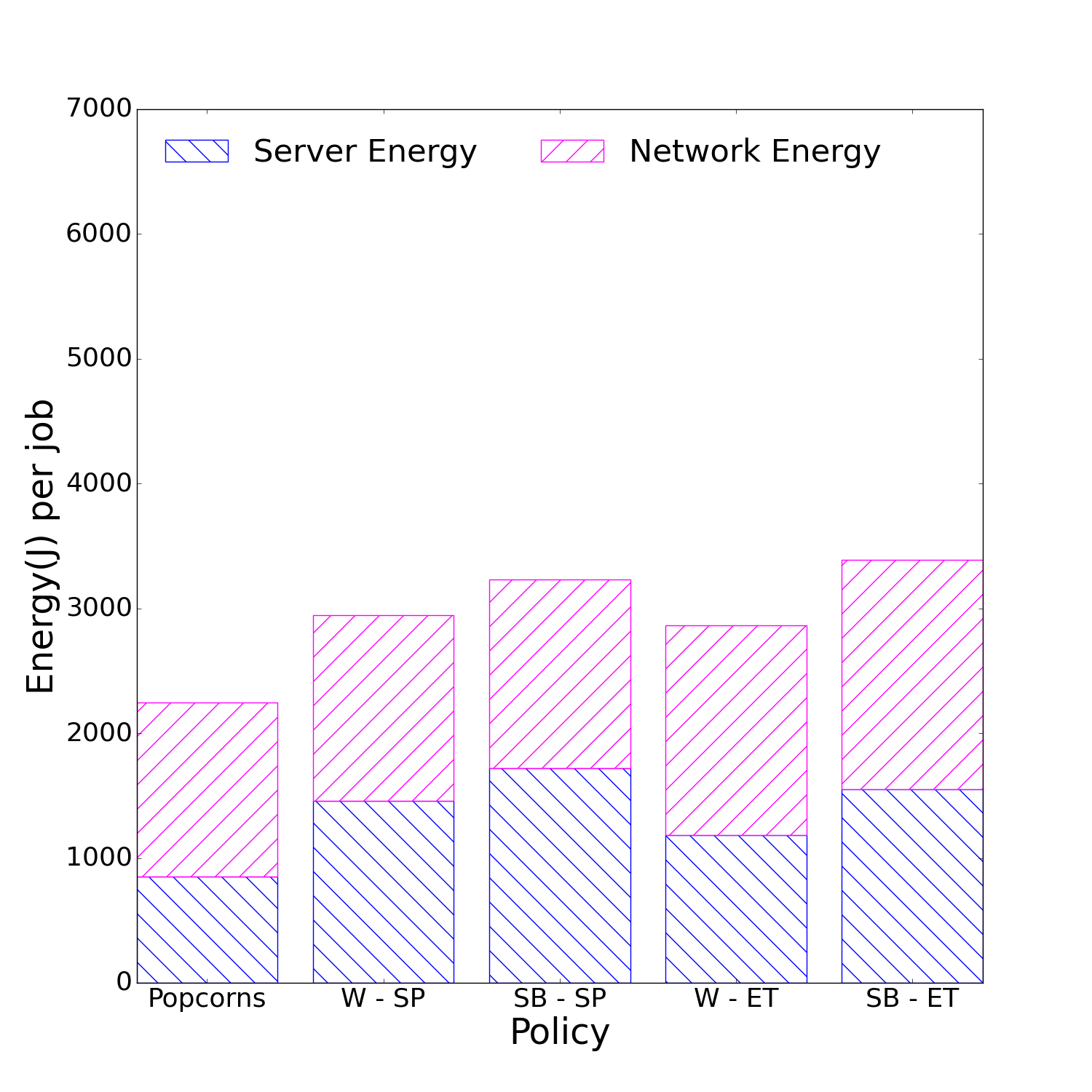}
	 	\label{elastic_tree_h_random}
	 }
	\subfloat[High Arrival Rate \newline $\lambda$=25 Jobs/Sec]{
		\includegraphics[width=0.33\textwidth]{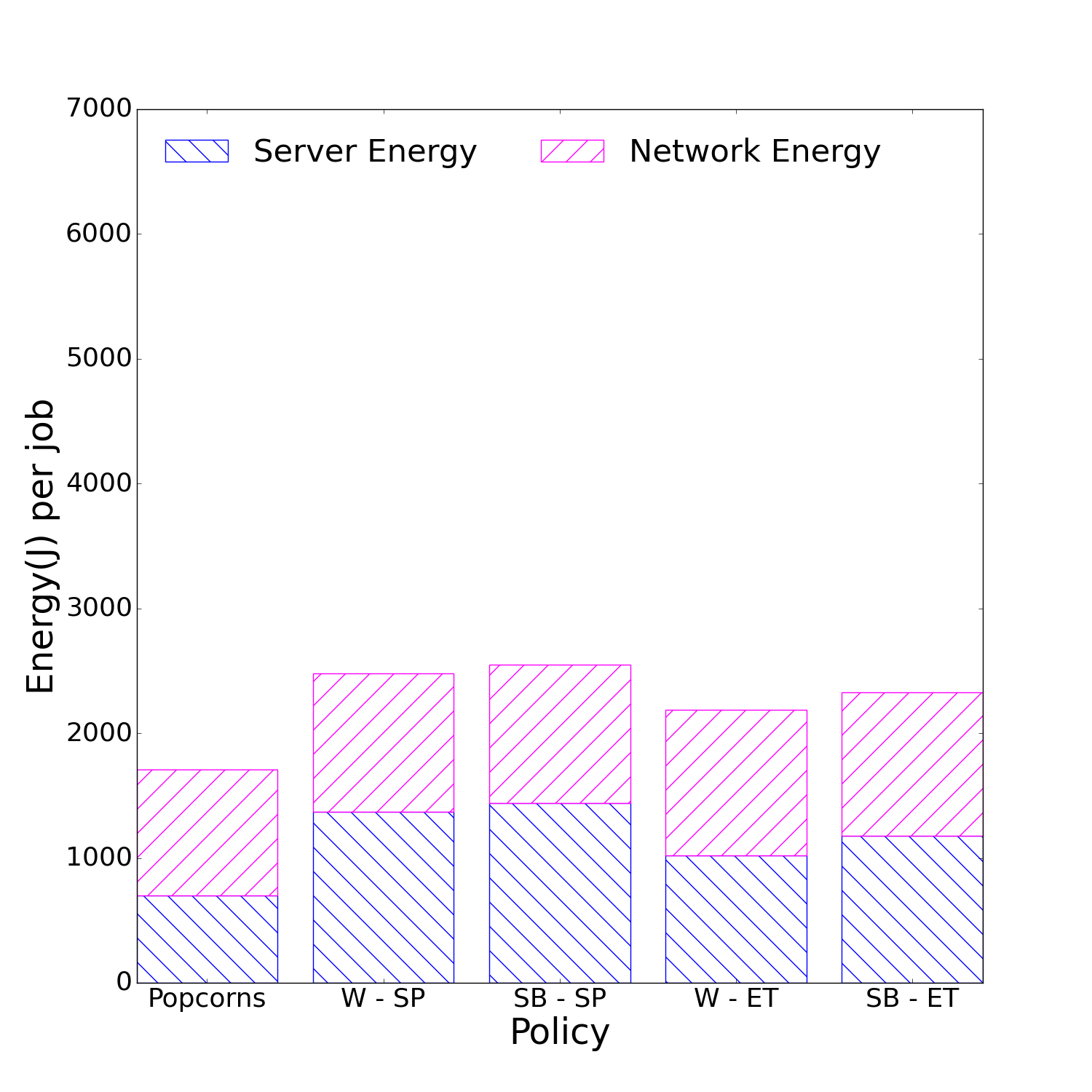}
		\label{elastic_tree_h_wasp}
	}
	
	\caption{Energy Savings comparison for different server and network path selection algorithms with MMPP based Job Arrival Pattern for search based 6-task job model. Every 30 seconds we have 10-second bursts with 1.5x Average Arrival rate.}
	\label{fig:mmpp-search}
	
\end{figure*}

\begin{figure*}[!h]
	\centering
	\captionsetup{font=small}
	\subfloat[Low Arrival Rate \newline Avg. Rate = 9 Jobs/Sec]{
		\includegraphics[width=0.33\textwidth]{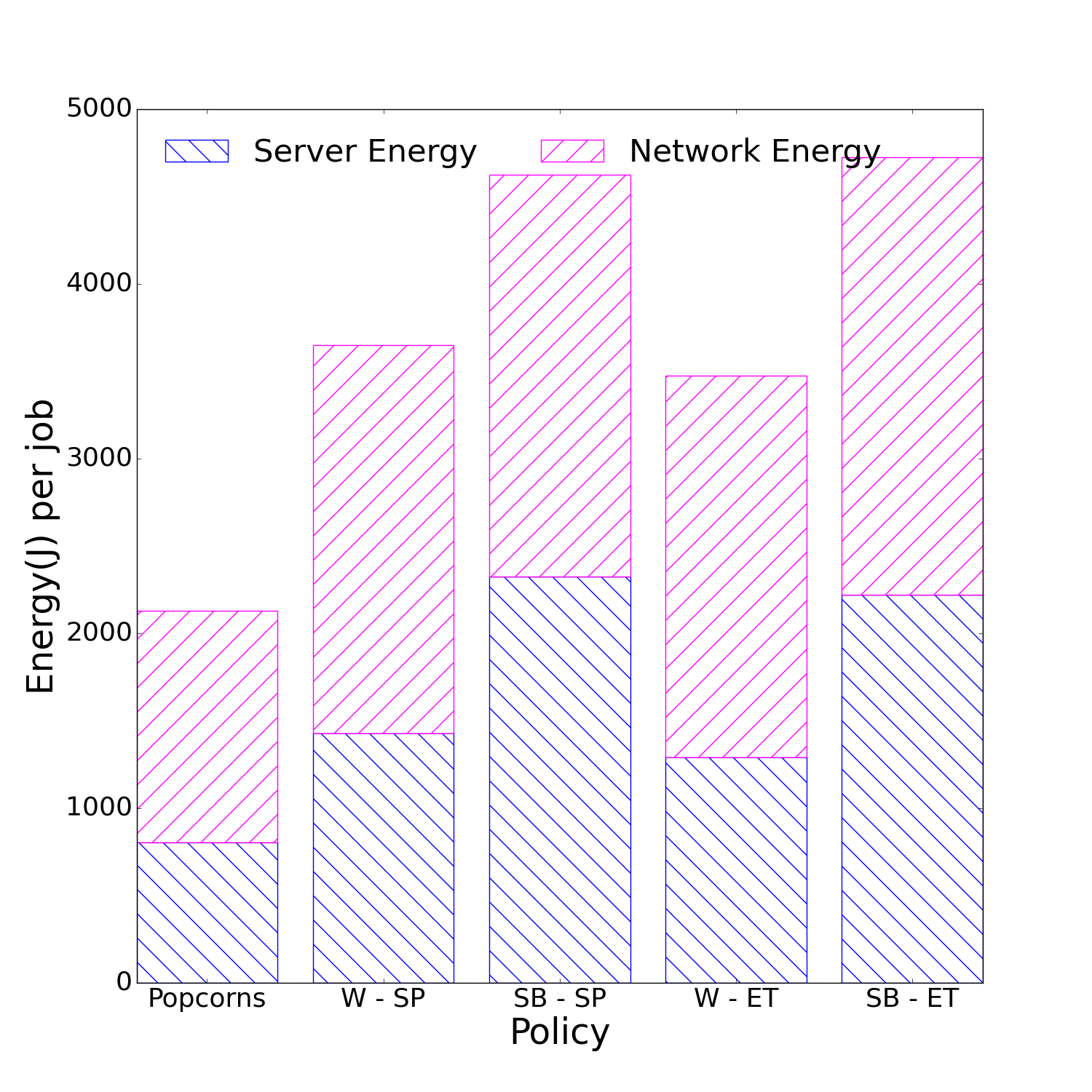}
		\label{djikstra_h_random}
	}
	 \subfloat[Medium Arrival Rate \newline Avg. Rate = 15 Jobs/Sec]{
	 	\includegraphics[width=0.33\textwidth]{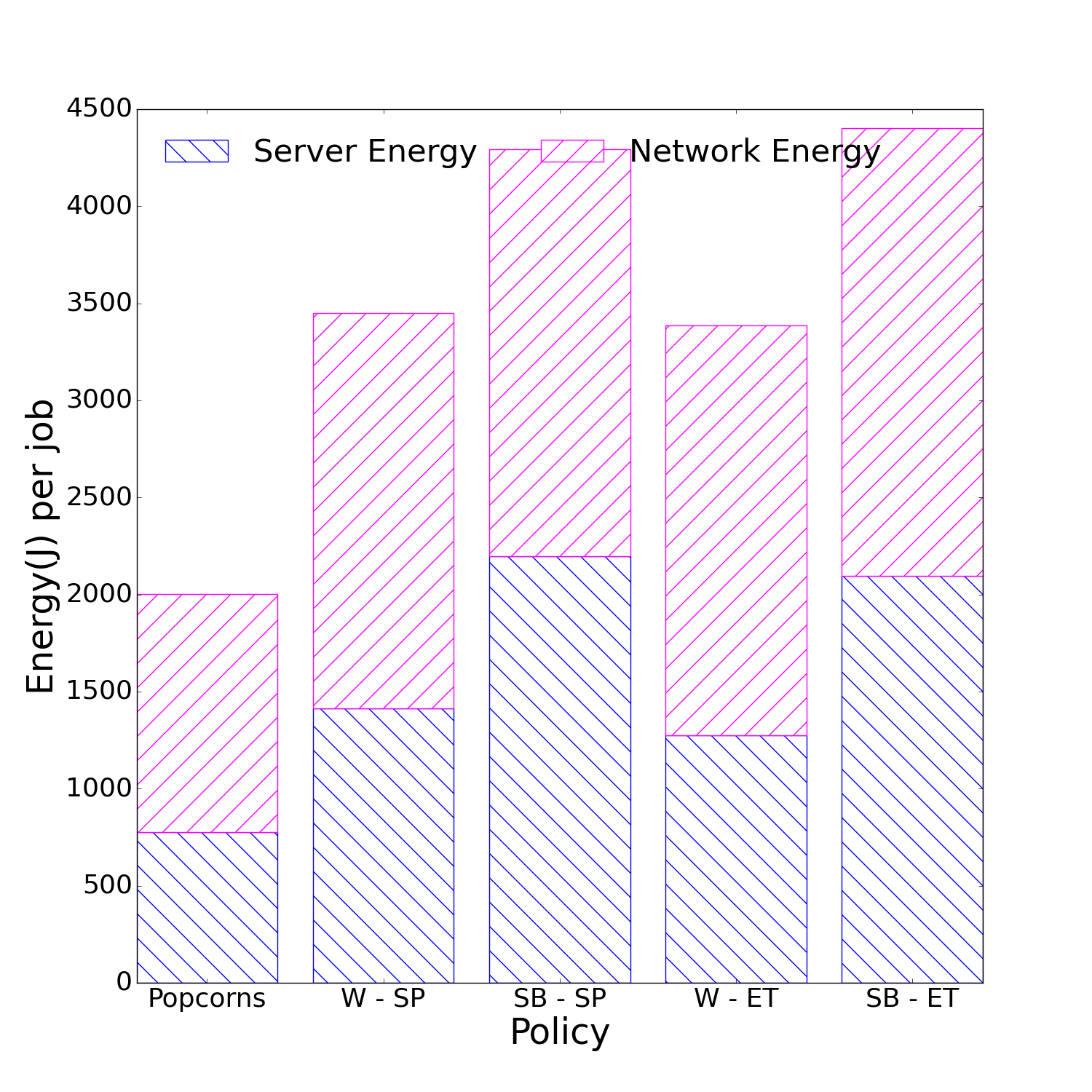}
	 	\label{elastic_tree_h_random}
	 }
	\subfloat[High Arrival Rate \newline Avg. Rate = 30 Jobs/Sec]{
		\includegraphics[width=0.33\textwidth]{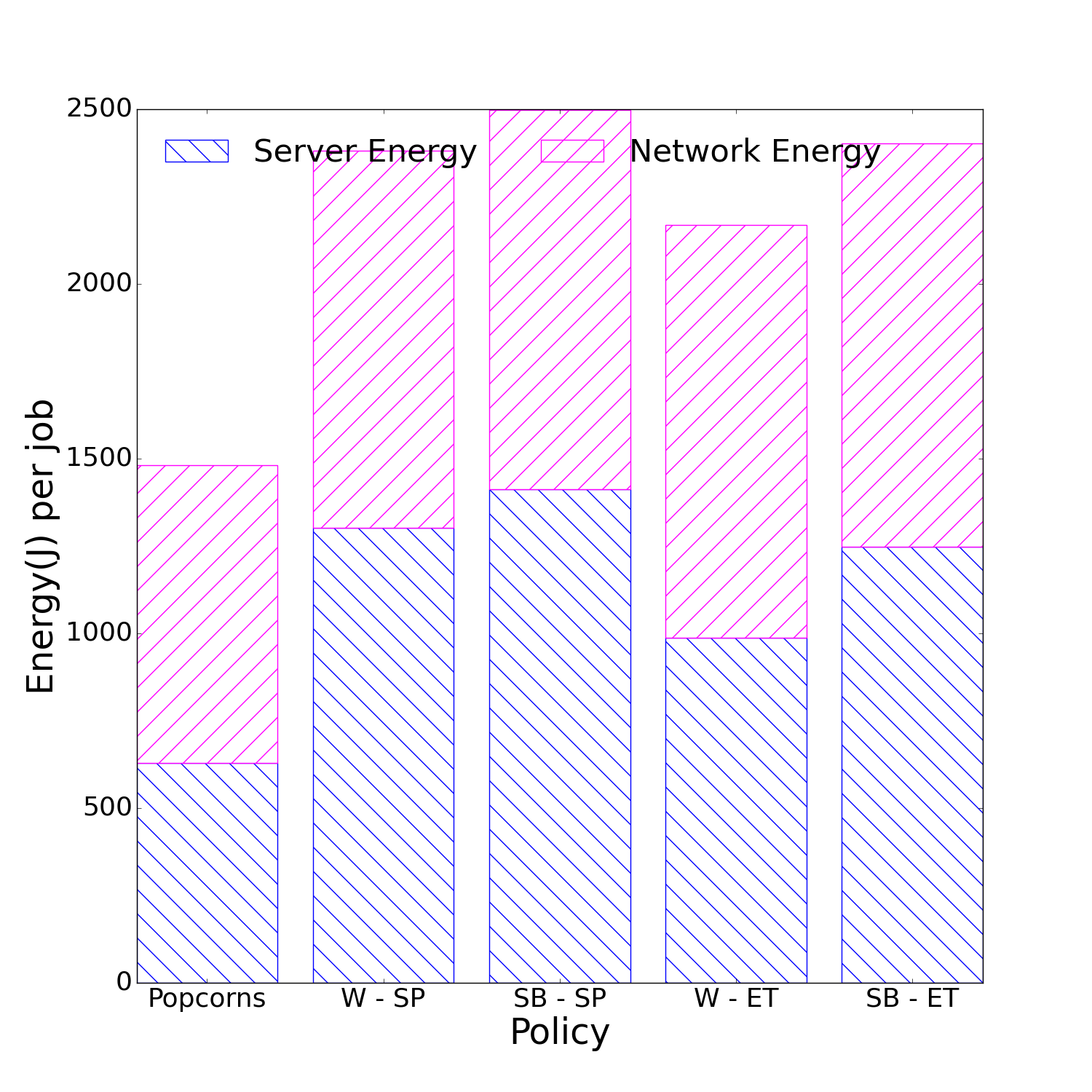}
		\label{elastic_tree_h_wasp}
	}
	
	\caption{Energy Savings comparison for different server and network path
    selection algorithms with NLANR trace based Job Arrival Pattern for search
    based 6-task job model.}
	\label{fig:trace-search}

\end{figure*}

\subsection{Evaluation Results}
\subsubsection{Energy Savings Comparison}
We compare the average energy consumption per job, for different combination of
network and server selection algorithm with Popcorn's selection algorithm. We
compare against a combination baseline policies WASP and random server
allocation policies with Elastictree and Shortest Path network selection
algorithms. As shown in Figure~\ref{fig:mmpp-ser}, popcorn algorithm for 2-task
service based job model for random arrival job provides significant savings
compared to Shortest-path based policies. This is attributed to fewer switches
and server being active. For ElasticTree based runs in figure
~\ref{fig:mmpp-ser}, using WASP as server allocation policy does provide more
opportunities for network consolidation and hence decreased savings. As we
increase the arrival rate to 60 Jobs/sec in figure, we see that Elastictree based runs perform
closer to \PopCorns as there are fewer opportunities for savings. We note that due to the bursty
nature of MMPP arrival rates (as discussed in section ~\ref{workloads}), the increased
delay in waking up additional switches leads to slower flow transmission rate
and increased energy consumption. We gather that server-only optimizations are
not always optimal when considering the whole data center energy consumption.
With NLANR trace based job arrivals. 

For Search based workload discussed in beginning of section ~\ref{sec:experiment}, the \PopCorns algorithm still results in significant savings.
We also see WASP based server optimization consistently performing better than Random server allocation. This is attributed to smaller communication size between each parent and child tasks and minimal time wasted waiting for the network data to arrive for processing.

\subsection{Flow size sensitivity}

\begin{figure}[h]
	\captionsetup{font=small}
	\centering\includegraphics[width=.85\linewidth]{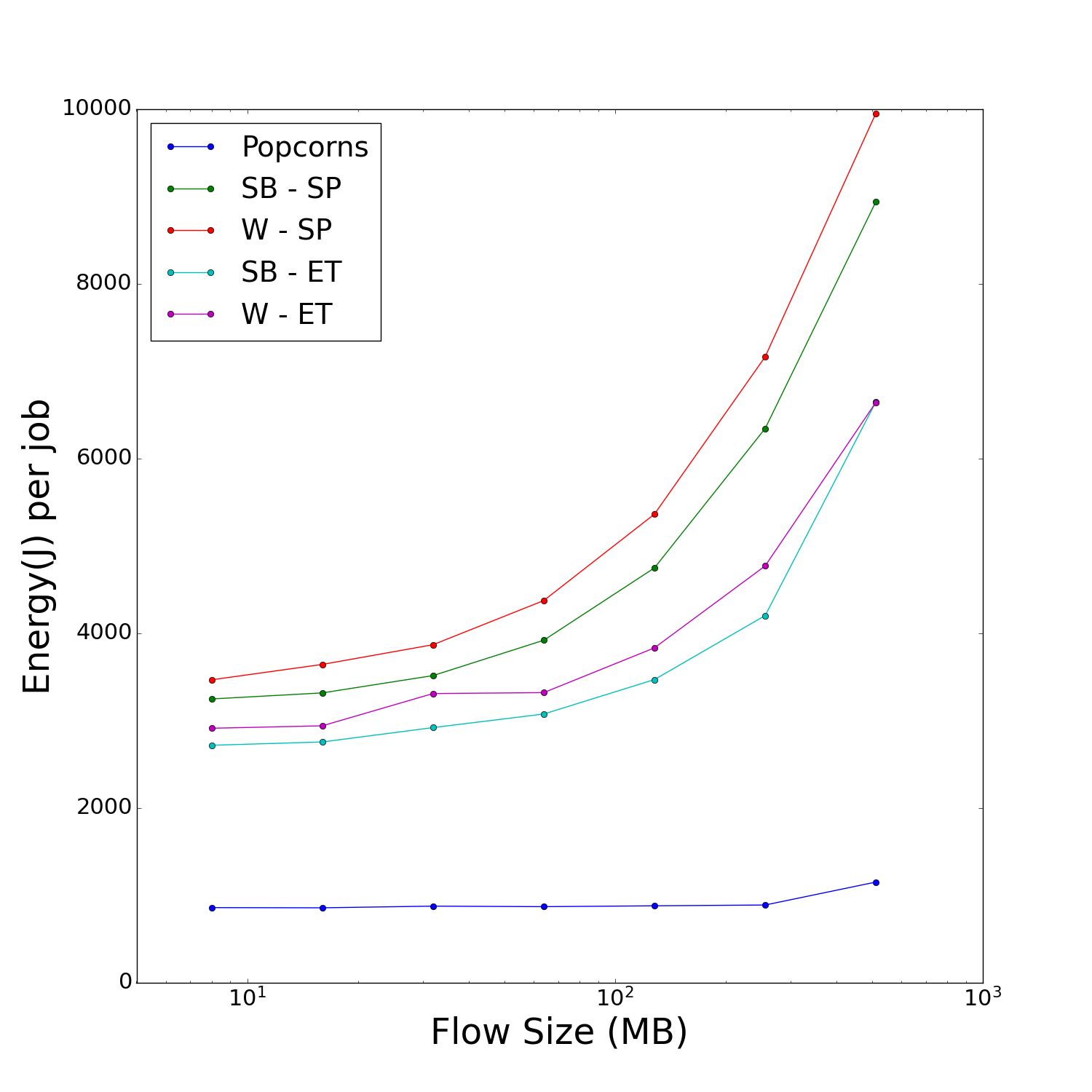}
	\caption{Energy consumption/Job for various flow sizes for different server
    scheduling and network routing algorithms. The experiment involves a 1024
    server fat-Tree topology with WebService jobs arriving randomly with a Poisson random
    variable value $\lambda$ = 25 jobs/sec and 10X QoS threshold.} 
	\label{fig:flows}
\end{figure}

The goal of this experiment is to determine if the routing path and server selection
by \PopCorns algorithm is resilient to increases in network flow sizes and
thereby increased network utilization. As we see in figure ~\ref{fig:flows} the
\PopCorns policy's is able to select paths and servers which are least affected
by the increase in flowsizes. Shortestpath based policies use most number of
switches and with increasing flow sizes there are fewer chances going to sleep
and hence consume the most energy. Elastic tree's greedy bin packing of flow
consolidation benefits from some network consolidation but with the non-optimal
server selection leading to longer flow communication times, the energy
consumption increases exponentially.

%

\subsection{Job completion latency Comparison}
\begin{figure}[!h]
	\captionsetup{font=small}
	\includegraphics[width=1\linewidth]{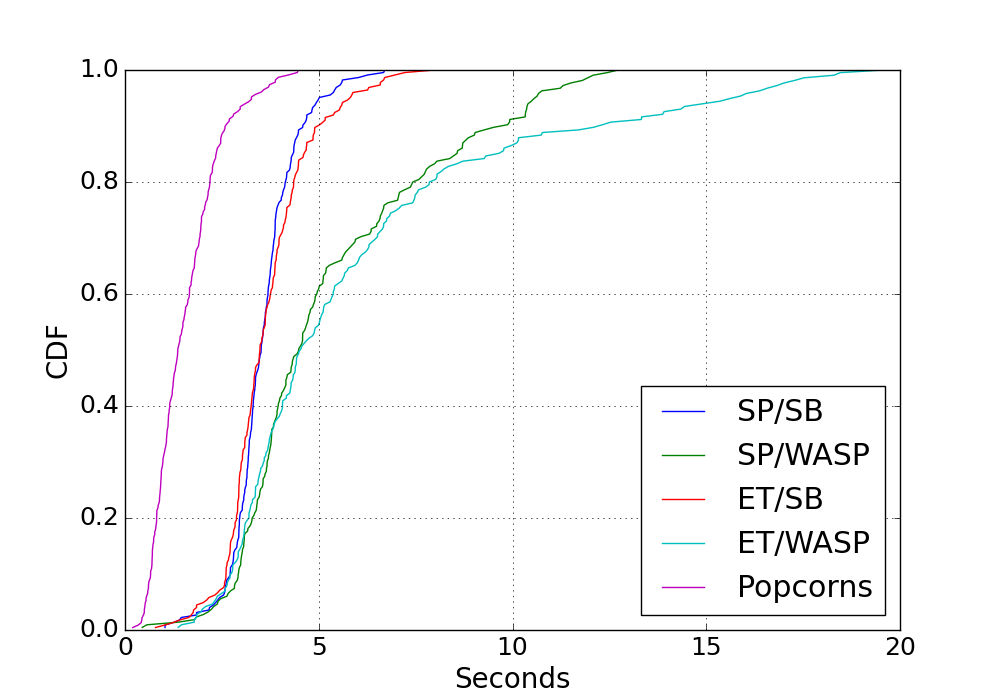}\caption{Cumulative-Distributed-function
    plot of job latencies for different server scheduling and network routing
    algorithms. The experiment involves a 1024 server fat-Tree topology with
    WebService modeled jobs arriving randomly with a Poisson random variable value $\lambda$ = 25
    jobs/sec and QoS threshold as 10X.}
	\label{cdf}
\end{figure}

Although all the algorithms discussed, consider the QoS latency threshold of
10x (Sum of Task Size of all the tasks in the job) for network
routing path selection, we see that the jobs in the \PopCorns algorithm are much
below the threshold in figure ~\ref{cdf}. Comparing WASP based policies with Server
balanced random server selection scheme, fewer servers being utilized in WASP,
the transmission time is higher and thus resulting in higher latencies.
\begin{figure} []
	\captionsetup{font=small}
	\includegraphics[width=1\linewidth]{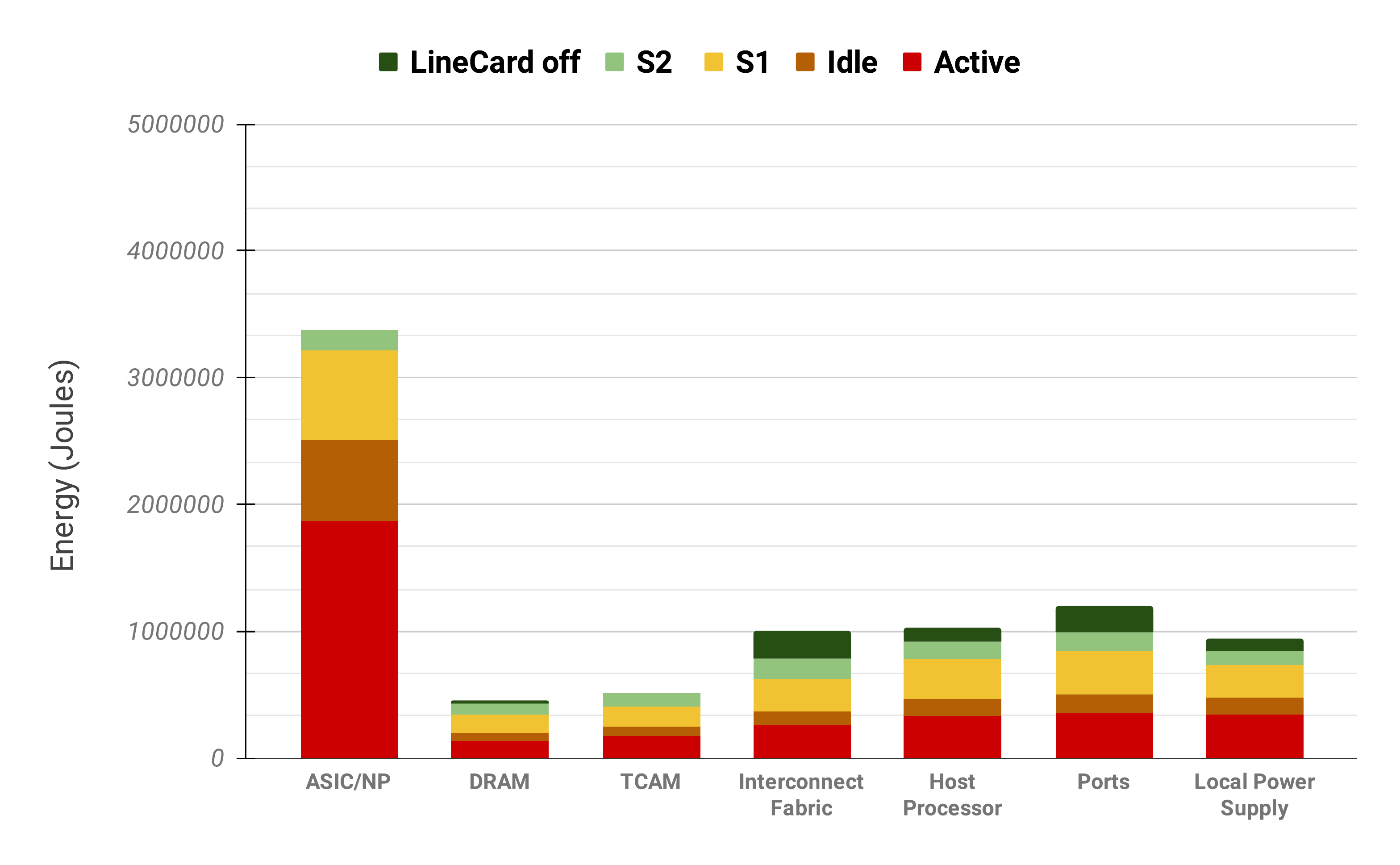}
	\caption{Component-wise energy distribution for Popcorns in Joules for 500
		second execution of 1024 server Fat-Tree topology with WebService model jobs arriving randomly (Poisson distribution) with $\lambda$ = 25 jobs per second. The energy consumed at different sleep states of the switch is also shown. }
	\label{fig:Popcorn-component}
\end{figure}

\begin{figure} []
	\captionsetup{font=small}
	\centering\includegraphics[width=1\linewidth]{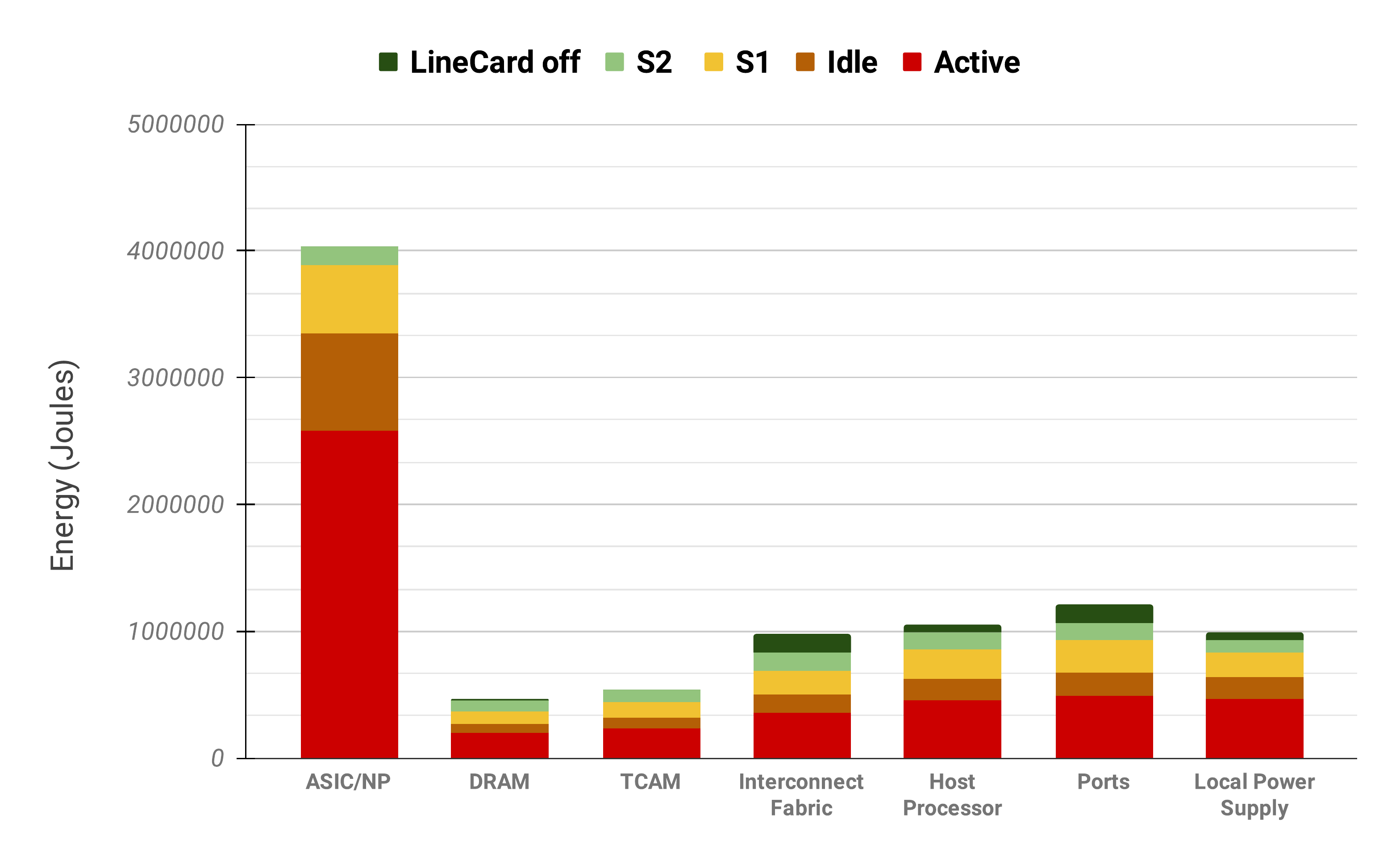}\caption{Component-wise
		energy distribution for WASP server scheduling policy with Elastic tree
		as network routing algorithm in Joules for 500 second execution of 1024
		server Fat-Tree topology with WebService model jobs arriving randomly (Poisson distribution) with $\lambda$=25 jobs per second. The energy consumed at different sleep states of the switch is also shown.}
	\label{fig:WASP-component}
	
\end{figure}

\subsection{Energy Distribution}

In this section we illustrate how system architects can utilize our simulated
settings and the power model to understand which component in the hardware
~\ref{fig:Popcorn-component} shows the energy consumed by each component in the
switch power model at each of the sleep state of the switch for typical
workload. This allows the system architect to focus their limited resources
towards optimizing a particular system architecture. For instance, looking at
figures ~\ref{fig:Popcorn-component} and ~\ref{fig:WASP-component}, we see that
most energy consumed by the network processor in its deepest sleep state. It can
also tell that a high cost design change for Sleep state S1 has more benefit
when using the Popcorns policy when compared to WASP and Elastictree policy.

%% file: baselines.tex
\begin{figure*}
	\centering
	\captionsetup{font=small}
	\subfloat[Shortest Path and Random Server job scheduling]{
		\includegraphics[width=0.33\textwidth]{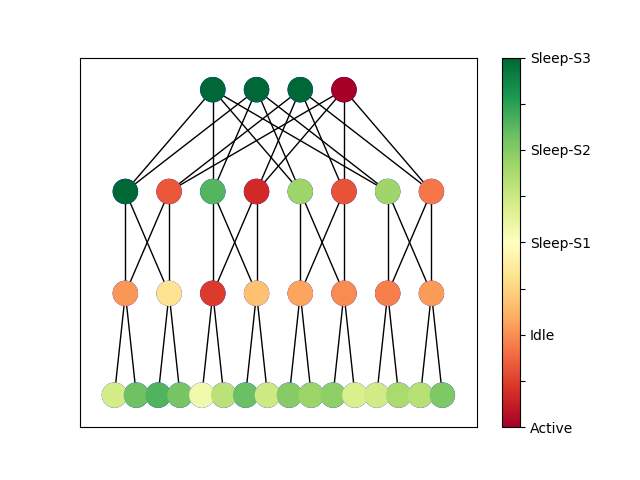}
		\label{djikstra_heat_random}
	}
	\subfloat[Elastic Tree and Random Server job scheduling]{
		\includegraphics[width=0.33\textwidth]{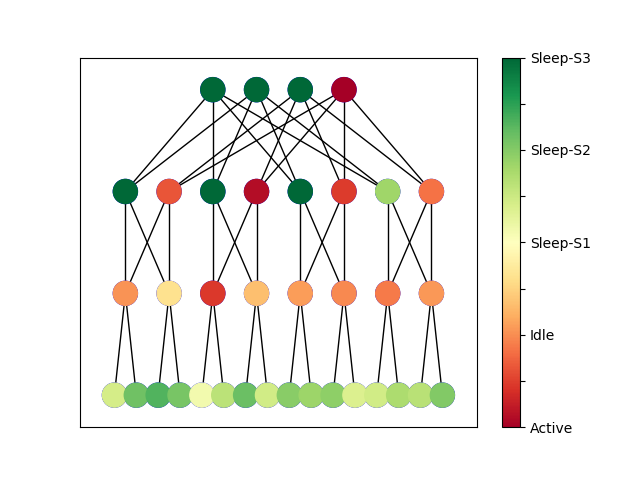}
		\label{elastic_tree_heat_random}
	}
	\subfloat[Shortest Path and WASP Server job scheduling]{
		\includegraphics[width=0.33\textwidth]{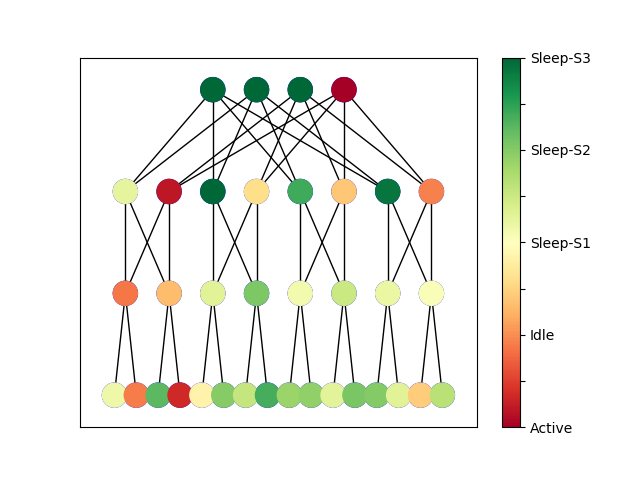}
		\label{djikstra_heat_wasp}
	}
	
	\subfloat[Elastic Tree and WASP Server job scheduling]{
		\includegraphics[width=0.33\textwidth]{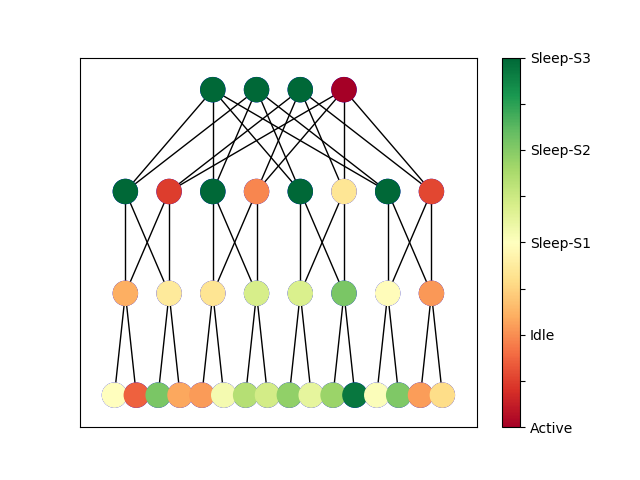}
		\label{elastic_tree_heat_wasp}
	}
	\subfloat[Popcorns based server and network scheduling]{
		\includegraphics[width=0.33\textwidth]{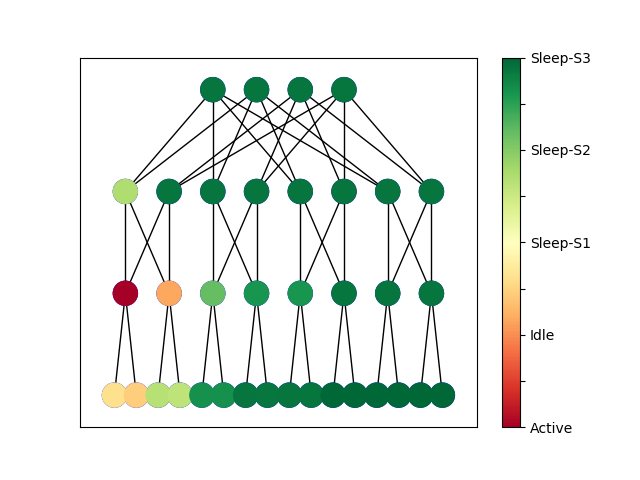}
		\label{popcorns_heat_popcorns}
	}
	\caption{Illustration of the average sleep state for switches and server with
		different policies. \PopCorns scheduling policy achieves greater consolidation of server and network flows. }
	\label{fig:heatmap-graph}
\end{figure*}

Figure ~\ref{fig:heatmap-graph}, illustrates the heat map after the end of the execution of the same Poisson based random job arrival arrival pattern with mean arrival rate $\lambda$=0.5 Jobs/second using the different server and network job scheduling policies for a small 16 server Fat-Tree topology. We note that with random server selection , all servers are equally utilized with elastic tree based network scheduling performing more network consolidation for better savings.
When using random server selection for tasks with Shortest path routing ~\ref{djikstra_heat_random} and Elastic based  routing ~\ref{elastic_tree_heat_random}, we can see that there are more aggregate switches in the dark green allowing more switches to stay in deep sleep state. Similarly for the WASP server selection scheme in figures ~\ref{djikstra_heat_wasp} 
and ~\ref{elastic_tree_heat_wasp}, we see that there are unequal usage of the servers and more flows consolidated into fewer switches. It can be see that even with server energy optimization in WASP, and our combined server and network policy \PopCorns ~\ref{popcorns_heat_popcorns} 
can achieve much higher savings than each of the individual optimizations. \PopCorns algorithm is able achieve this mainly due to being aware of network energy aware state when scheduling jobs on the servers. It is further able to reduce energy savings by increased consolidation of flows and servers by utilizing the latency slacks available at each node in the data center.

%% file: RelatedWork.tex
\section{Related Work}
\label{sec:RelatedWork}
With the energy consumption of large data centers reaching Gigawatt scale, its energy saving techniques are increasingly being studied in recent years. 
Common techniques used for server energy reduction include Dynamic voltage and
frequency scaling (DVFS) to reduce the energy at the cost of server performance
 ,Co-ordinated DVFS and sleep states for
server processors \cite{yaoTSBatProImprovingEnergy2018} \cite{yaoWASPWorkloadAdaptive2017}, and
virtualization to consolidate VMs into fewer servers
\cite{hsuOptimizingEnergyConsumption2014}.  TS-Bat ~\cite{yaoTSBatProImprovingEnergy2018}
demonstrates that, through temporally batching the jobs and by grouping them
onto specific servers spatially, higher power savings can be obtained.
WASP~\cite{yaoWASPWorkloadAdaptive2017} shows that intelligent use of low power states in
servers can be used to boost server power savings.

For the energy efficiency in network, earlier works have looked at switches and
routers for Internet-scale large area networking. Gupta et
al.~\cite{guptaGreeningInternet2003} first proposed the need for power saving in
networks and pointed to having network protocol support for energy management.
Adaptive Link Rate (ALR) \cite{gunaratneReducingEnergyConsumption2008} reduces
link energy consumption by dynamically adjusting data link rate depending on
traffic requirements. Other approaches include turning off switches when not
required, or to put them in sleep mode depending on packet queue length
\cite{yuEnergyefficientDelayawarepacket2015}
 Prior work on reducing data center network
power rely on DVFS and sleep states \cite{iqbalEfficienttrafficaware2012} to
opportunistically reduce power consumption of individual switches. In these
approaches, switches may enter sleep states without knowledge of incoming server
traffic and may be forced to wake up prematurely. We study a co-ordinate server
job placement and network allocation required to optimize the amount of sleep
time and save network energy consumption.Recently,
DREAM~\cite{zhouDREAMDistRibutedEnergyAware2019} proposed a probability based
network traffic distribution scheme by splitting flows to save network power.

There are existing works which combine server and network power saving.
Mahadevan et al.\cite{mahadevanEnergyAwareNetwork2009} and Heller et al.
~\cite{hellerElasticTreeSavingEnergy2010} have proposed a heuristic based
algorithm for a coarse-grained load variation which dynamically allocates the
servers required for the workload and powers off the unneeded switches for the
server configuration. Other approaches consolidate VMs in fewer servers and in
turn fewer switches \cite{zhengJointpoweroptimization2014}. These approaches
assume an unrealistically high amount of idle period to offset the large wakeup
latencies to transition between On and Off states. To the best of our knowledge
our solution is the only one to consider network sleep states to target higher
power savings in the data center. The EEPRON~\cite{zhouJointServerNetwork2018}
discussed using the network slack available to reduce the energy consumption of
servers with frequency scaling.


%% file: Conclusion.tex
\section{Conclusion}
\label{sec:Conclusion}
In this article, we presented \PopCorns, where we explore techniques that make smart use of line card and port low power states in switches and orchestrate them with intelligent joint task placement and routing algorithm for more effective power management. The results show good promise in achieving considerable power savings compared to the baseline policies. Our experimental results show that smart management of low power states achieves upto 80\% over policies optimizing servers and network policies separately, while still keeping energy savings low.

%% file: acknowledgement.tex
\section*{Acknowledgment}
\label{sec:acknowledgement}
This material is based in part upon work supported by the National Science Foundation under Grant Number CNS-178133. Kathy Nguyen and Gregory Kahl were supported through a REU supplement under the NSF award.